\documentclass[peprint,3p,times]{elsarticle}
\usepackage[utf8]{inputenc}
\usepackage{color,url,bbm,mathtools}
\usepackage{xcolor}
\usepackage{colortbl}
\usepackage{booktabs}
\usepackage{multirow}
\usepackage{multicol}
\usepackage{verbatim}
\usepackage[normalem]{ulem}
\usepackage{bm}
\usepackage{layout}
\usepackage{subcaption}
\usepackage{cancel}
\numberwithin{equation}{section}
\numberwithin{figure}{section}
\usepackage{amsfonts}
\usepackage{amsmath}
\usepackage{amssymb}
\usepackage{amstext}
\usepackage{amsthm}
\usepackage{cancel}
\usepackage{comment}
\usepackage{systeme}
\usepackage{algorithm}
\usepackage{algpseudocode}
\usepackage{makecell}
\usepackage{adjustbox}
\usepackage[colorlinks=true,urlcolor=blue,linkcolor=blue,citecolor=blue]{hyperref}
\usepackage[capitalise,noabbrev]{cleveref}
\crefformat{equation}{(#2#1#3)}
\usepackage{mdframed}

\usepackage{tikz}
 
\usepackage{float, caption, subcaption, graphicx}
\usepackage[group-separator={,},group-minimum-digits={3}]{siunitx}

\newcommand{\romcoef}[1]{{a_{#1}}}
\newcommand{\oa}{\overline{a}}
\newcommand{\oua}{\overline{\ua}}
\newcommand{\romvec}{\ua}
\newcommand{\filtercoef}[1]{{\oa_{#1}}}
\newcommand{\filtervec}{\oua}

\newcommand{\mF}{\mathcal{F}}
\newcommand{\mL}{\mathcal{L}}
\newcommand{\uwb}{\widetilde{\ub}}

\newcommand{\dlrom}{\text{StabOp-L-ROM}}
\newcommand{\fom}{\text{FOM}}
\newcommand{\rom}{\text{ROM}}
\newcommand{\qoi}{\text{QoI}}
\newcommand{\Ttrain}{\mathcal{T}_{\text{train}}}
\newcommand{\Tval}{\mathcal{T}_{\text{val}}}
\newcommand{\Ttest}{\mathcal{T}_{\text{test}}}

\newcommand{\real}{\mathbb{R}}

\newcommand{\tinit}{{t}_{\text{init}}}
\newcommand{\ttrain}{{t}_{\text{train}}}
\newcommand{\tval}{{t}_{\text{val}}}
\newcommand{\tfinal}{{t}_{\text{final}}}

\newcommand{\EFOM}{E_{\text{FOM}}}
\newcommand{\EROM}{E_{\text{ROM}}}
\newcommand{\Ntrain}{{N}_{\text{train}}}
\newcommand{\Nval}{{N}_{\text{val}}}
\newcommand{\egrowth}{{\epsilon}_{\text{growth}}}
\newcommand{\valmetric}{\mathcal{E}_{\text{val}}}
\newcommand{\tA}{\widetilde{A}}
\newcommand{\tB}{\widetilde{B}}
\newcommand{\tb}{\widetilde{b}}

\newcommand{\Nepoch}{{N}_{\text{epochs}}}
\newcommand{\mP}{\mathcal{P}}

\def\scriptO{{{\it O}\kern -.42em {\it `}\kern + .20em}}
\def\RR{{{\rm l}\kern - .15em {\rm R} }}
\def\PP{{{\rm l}\kern - .15em {\rm P} }}
\def\VV{{{\rm V}\kern - .69em {\rm V} }}
\def\L2{{{\sf L}^2}}
\def\H1{{{\sf H}^1}}
\def\PN2{{\PP_{N}-\PP_{N-2}}}
\def\QM2{{\mathcal{Q}_{M}-\mathcal{Q}_{M-2}}}

\def\complex{{{\rm C} \kern - .53em {\rm l} \kern + .38em}}

\def\a1{{ | \lambda_{\min} |}}

\def\l1{{   \lambda_{\min}  }}

\def\bff{{\bf f}}

\def\bu0{{\underline {\bf 0}}}
\def\buu{{\underline {\bf u}}}

\def\bu{{\bm u}}

\def\bv{{\bf v}}
\def\bx{{\bf x}}

\def\bX{{\bf X}}

\def\ua{{\boldsymbol a}}
\def\ub{{\boldsymbol b}}

\def\u0{{\underline 0}}

\def\bX{{\bf X}}

\def\utheta{{\bf{\underline \theta}}}

\def\tr{\text{r}}

\def\obu{\overline{\bu}}

\def\utheta{\boldsymbol{\theta}}

\newtheorem{remark}{Remark}[section]

\def\PP{{{\rm l}\kern - .15em {\rm P} }}
\def\PN2{{\PP_{N}-\PP_{N-2}}}

\newcommand{\cF}{\mathcal{F}}

\newcommand{\bphi}{\boldsymbol{\varphi}}

\definecolor{vargreen}{rgb}{0.0, 0.5, 0.0}

\newcommand{\deleted}[1]{{}}

\begin{document}

\begin{frontmatter}

\title{
StabOp: A Data-Driven Stabilization Operator 
for Reduced Order Modeling}

\author[1]{Ping-Hsuan Tsai \corref{cor1}}
\ead{pinghsuan@vt.edu}

\author[2]{Anna Ivagnes}
\ead{aivagnes@sissa.it}

\author[3]{Annalisa Quaini}
\ead{aquaini@central.uh.edu}

\author[1]{Traian Iliescu}
\ead{iliescu@vt.edu}

\author[2]{Gianluigi Rozza}
\ead{grozza@sissa.it}

\cortext[cor1]{Corresponding author}

\affiliation[1]{organization={Department of Mathematics, Virginia Tech},
            city={Blacksburg},
            postcode={24061}, 
            state={VA},
            country={United States}}
            
\affiliation[2]{organization={Mathematics Area, mathLab, SISSA, International School for Advanced Studies},
            city={Trieste},
            postcode={34136}, 
            country={Italy}}
\affiliation[3]{organization={Department of Mathematics, University of Houston},
city={Houston}, postcode={77204}, state={TX}, country={United States}}

\begin{abstract}
Spatial filters have played a central role in large eddy simulation for many decades and, more recently, in reduced order model (ROM) stabilization for convection-dominated flows. Nevertheless, important open questions remain: In under-resolved regimes, which filter is most suitable for a given stabilization or closure model? Moreover, once a filter is selected, how should its parameters, such as the filter radius, be determined? Addressing these questions is essential for the reliable design and performance of filter-based stabilization or closure strategies.
To answer these critical questions, we propose a novel strategy that is fundamentally different from current filter-based stabilizations and closures:
We replace the traditional spatial filters with a novel data-driven stabilization operator (StabOp) that yields the most accurate results for a given resolution, quantity of interest, and stabilization strategy.  
Although the new StabOp could be used for both classical discretizations and ROMs, and for different types of filter-based stabilization or closure, for clarity, we investigate it for ROMs and the Leray ROM (L-ROM) stabilization.
To build the new StabOp, we first postulate its model form to be a linear mapping, a quadratic mapping, or a nonlinear mapping (through a neural network), and then solve a PDE-constrained optimization problem to minimize the given loss function. 
Using the resulting StabOp in the given L-ROM yields a new stabilized ROM, StabOp-L-ROM. 
To assess the new StabOp-L-ROM, we compare it with the L-ROM and the standard ROM in the numerical simulation of four convection-dominated flows in the under-resolved regime: 
2D flow past a cylinder at $\rm Re=500$, lid-driven cavity at $\rm Re=10000$, 3D flow past a hemisphere at $\rm Re=2200$, and minimal channel flow at $\rm Re=5000$. 
Our numerical results demonstrate that the new StabOp-L-ROM can be orders-of-magnitude more accurate than the classical L-ROM tuned with the optimal filter radius in the predictive regime. Furthermore, while the new StabOp smooths the input flow fields, its smoothing mechanism is entirely different from those of classical spatial filters.

\end{abstract}


\end{frontmatter}

\section{Introduction}

{\it Spatial filters} have made a profound impact in the numerical simulation of convection-dominated flows. 
In full order models (FOMs), i.e., computational models obtained by using classical numerical methods (e.g., the finite element method), spatial filters have been the main tool used to develop {\it large eddy simulation (LES)} models. See the research monographs~\cite{BIL05,garnier2009large,Pop00,rebollo2014mathematical,sagaut2006large}.
Indeed, in turbulent flow simulations, the number of degrees of freedom required by a direct numerical simulation (i.e., a resolved simulation) to capture all the spatial scales in the flow is prohibitively high.
LES addresses this important practical issue by leveraging spatial filters:
First, the underlying equations are filtered with a given spatial filter to eliminate the small scales and keep only the large spatial scales that can be represented on the given coarse mesh.
Then, the equations for the large scales (i.e., the LES model) are solved to approximate the large spatial structures in the flow.
Since the underlying equations are nonlinear, the LES model is not closed and, thus, the closure problem must be addressed. 
That is, a model for the interaction between the large, resolved scales and the small, unresolved scales must be constructed.  
LES models have been central in the numerical simulation of turbulent flows:
There are research monographs devoted to LES for both incompressible \cite{BIL05,Pop00,rebollo2014mathematical,sagaut2006large} and compressible flows \cite{garnier2009large}, and LES models are available in widely used software (e.g., ANSYS Fluent~\cite{manual2009ansys},
COMSOL Multiphysics \cite{Comsol}, 
Simcenter STAR-CCM+ \cite{Simcenter}, 
Nek5000~\cite{fischer2008nek5000}, NekRS~\cite{fischer2022nekrs}, etc).

More recently, spatial filters have also made a significant impact in reduced order models (ROMs), which are efficient alternatives to FOMs. Just as for FOMs, in the numerical simulation of convection-dominated (e.g., turbulent) flows, the relatively low-dimensional ROMs often yield inaccurate solutions, typically manifested as spurious numerical oscillations.
To mitigate these inaccuracies in the realistic under-resolved regime, two types of strategies are generally used: 
(i) {\it ROM closures} (see the review in \cite{ahmed2021closures}), which supplement the ROM with correction terms to model the influence of the small, unresolved spatial scales on the large, resolved scales; and
(ii) {\it ROM stabilizations} (see the review in \cite{parish2024residual}), which modify existing ROM terms or add new ones to enhance the numerical stability.
We note that, because the primary role of the unresolved small scales is to dissipate energy in the reduced system~\cite{CSB03}, ROM closures and stabilizations often overlap.
We also emphasize that ROM spatial filters (e.g., the ROM projection, ROM differential filter, and ROM higher-order algebraic filter \cite{tsai2025time}) have been central in the construction of both ROM closures and ROM stabilizations. 
We also note that spatial filtering has also been used to filter the input data \cite{aradag2011filtered,farcas2022filtering} to increase the ROM stability and accuracy.

To build ROM closures based on spatial filtering, the LES framework has been used as a starting point.
Specifically, ROM spatial filters were first leveraged to determine the ROM closure term, which was then approximated through various approaches.
Examples of ROM closures include the approximate deconvolution ROM \cite{xie2017approximate} and the variational multiscale ROM \cite{baiges2015reduced,MANTI2025114298,mou2021data,reyes2020projection}.
Many more examples of ROM closures are surveyed in \cite{ahmed2021closures}.

To build ROM stabilizations based on spatial filtering (see the review in \cite{quaini2024bridging}), ROM spatial filtering is used to smooth out either selected terms or all terms in the standard ROM. 
The principle underlying filter-based ROM stabilizations can be summarized as follows:
Apply ROM spatial filtering to smooth (regularize) specific ROM terms or the entire ROM solution, with the objective of increasing ROM stability and thereby improving accuracy. Examples of filter-based ROM stabilizations include the Leray ROM (L-ROM) \cite{tsai2022parametric,wells2017evolve}, the evolve-filter-relax ROM \cite{girfoglio2021pod,girfoglio2023linear,gunzburger2019evolve, ivagnes2025data,strazzullo2022consistency,wells2017evolve}, and the recently introduced time relaxation ROM \cite{tsai2025time}. These filter-based ROM stabilizations have been employed in challenging convection-dominated flows, including flow past a cylinder~\cite{girfoglio2021pod,ivagnes2025data}, 
lid-driven cavity flow~\cite{kaneko2020towards}, quasi-geostrophic dynamics~\cite{girfoglio2023linear}, and turbulent channel flow~\cite{tsai2025time}, and they have shown success in producing efficient and accurate simulations.
They have also been implemented in 
software, e.g., NekROM~\cite{kaneko_nekrom} and
ITHACA-FV \cite{ITHACAFV}.

Finally, we also note that spatial filtering has also been recently leveraged to increase the stability of neural-network-based models~\cite{rezaian2023predictive}.

\begin{figure}[!ht]
    \includegraphics[width=1\textwidth]{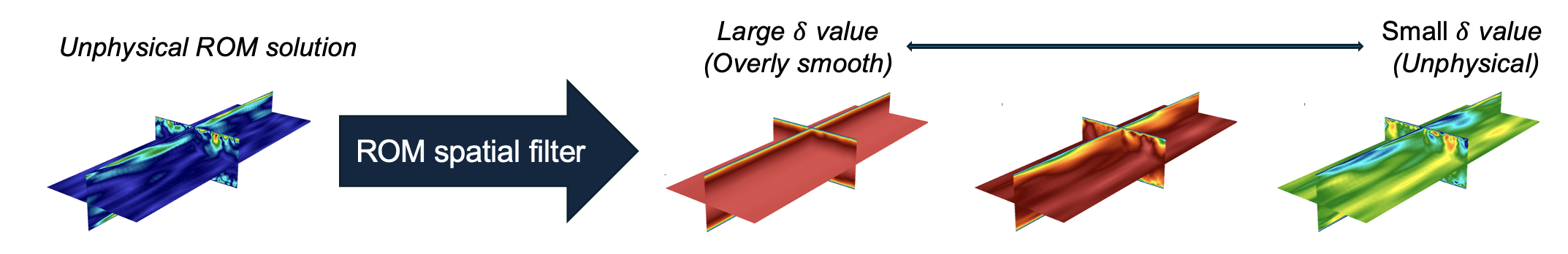}
   \caption{Effect of the ROM spatial filter in 
   3D turbulent channel flows \cite{tsai2025time}. A suitably chosen filter radius $\delta$ yields a physically meaningful solution, while small or large $\delta$ values lead to unphysical or overly smooth results, respectively.}
   \label{fig:filtered-config}
\end{figure}

Despite the undeniable success of spatial filtering in computational fluid dynamics, there are still open questions. 
Indeed, for a given under-resolved setting (dictated by the limited available computational resources) and a given LES model or filter-based ROM stabilization, consider the following {\it practical} questions:
(i) {\it Which spatial filter should we choose?}
At a FOM level, we could choose from, e.g., linear filters (such as the Gaussian, spectral cutoff, box, or differential filters)~\cite{sagaut2006large} and nonlinear filters \cite{layton2012approximate}.
At a ROM level, we could choose from the ROM projection, the ROM differential filter, or the higher-order algebraic filter.  
We emphasize that the answer to this question can be critical to the success of the chosen model: different types of filters might lead to different accuracy levels.
(ii) Another important practical question is:
{\it For a chosen filter, how do we choose its parameters?}
For example, if the differential filter is chosen to build an LES model, how do we choose the {\it filter radius}?
Again, this question is important in practice, e.g., when using the ROM differential filter to build the stabilized model \cite{mou2023energy}, as illustrated in \cref{fig:filtered-config}: 
A suitably chosen filter radius, $\delta$, yields a physically meaningful solution. 
In contrast, too small $\delta$ values produce unphysical oscillations, and too large $\delta$ values lead to an overly smoothed result. 

In this paper, we propose a novel strategy to answer the above questions:
Instead of using traditional spatial filters, we propose a {\it data-driven stabilization operator (\textbf{StabOp})} that yields the most accurate results for a given resolution, a given quantity of interest (QoI), and a given type of stabilization strategy.
Before outlining the StabOp's construction, we emphasize that it is {\it fundamentally different from standard spatial filters}.
Indeed, rather than \emph{assuming} that one of the classical spatial filters is the optimal choice for constructing a given stabilization, we turn the problem on its head: 
{\it We use a data-driven strategy to identify the operator (i.e., StabOp) that produces the most accurate stabilization.} 

To illustrate the construction of the novel StabOp, we choose a specific filter-based stabilization, the Leray model.
Furthermore, for simplicity of presentation, we illustrate the StabOp's construction for ROMs.
We emphasize, however, that the new {\it StabOp strategy is general} and could be used both for FOMs and for ROMs, and for  different types of filter-based stabilization and closures.
To build the new StabOp, we first consider it as a general ROM operator acting from the ROM space to itself, viewed as a mapping from input, unstabilized ROM coefficients to output, stabilized ROM coefficients.
We then postulate a model form for StabOp, such as {\it a linear mapping, a quadratic mapping, or a nonlinear mapping defined by a neural network}, and determine the StabOp's parameters by solving a {\it PDE-constrained optimization} problem that minimizes an appropriately defined loss function. 
This loss depends on the given QoI and may quantify, for example, the discrepancy between the kinetic energy of the resulting data-driven stabilized ROM (i.e., in our case, the Leray ROM with StabOp stabilization, which we denote as \textbf{StabOp-L-ROM}) and that of the FOM.
In the training regime, by construction, the new StabOp-L-ROM yields more accurate results than L-ROM equipped with any spatial filter.
A natural question is whether the new {\it StabOp-L-ROM is also 
accurate in the predictive regime}. Our numerical investigation shows that this is indeed the case. 
Another natural question is how the new StabOp compares with classical spatial filters, such as the ROM differential filter and the ROM projection.
Our numerical investigation shows that, while the new StabOp generally smooths the input flow fields (like classical spatial filters), it yields results that are different from those obtained by applying 
standard ROM spatial filters, e.g., the ROM differential filter and the ROM projection.

The rest of the paper is organized as follows:
In Section~\ref{section:numerical-modeling}, we outline the FOM used to generate the snapshots and the standard Galerkin ROM (G-ROM). 
Section \ref{sec:les-rom} reviews the Leray ROM (L-ROM), which is the filter-based ROM stabilization chosen to illustrate the new StabOp's construction. 
In Section~\ref{section:d2-les-rom}, we present a general formulation for the new StabOp-L-ROM (i.e., the ROM stabilization obtained by combining the Leray ROM with the new StabOp) and the computational implementation of the data-driven strategy.  
In Section~\ref{section:numerical-method}, we outline the computational implementation of the new StabOp and the resulting ROM stabilization, StabOp-L-ROM.
In Section~\ref{section:numerical-results}, we present a numerical investigation of the new StabOp-L-ROM, comparing it against the classical L-ROM in which the filter radius is optimally chosen, and the standard G-ROM. 
In our numerical investigation, we consider the following convection-dominated flows: the 2D flow past a circular cylinder at $\rm Re=500$, the 2D lid-driven cavity at $\rm Re=10000$, the 3D flow past a hemisphere at $\rm Re=2200$, and the 3D minimal channel flow at $\rm Re=5000$. Finally, in Section \ref{section:conclusions}, we summarize our findings and outline directions for future work. 

\section{Numerical Models}
    \label{section:numerical-modeling}

In this section, we briefly outline the FOM (Section~\ref{section:fom}) and the
G-ROM (Section~\ref{section:g-rom}) used in our numerical investigation.

\subsection{Full Order Model (FOM)}
    \label{section:fom}
    
As a mathematical model, we consider the incompressible Navier-Stokes equations (NSE) with forcing: 
\begin{align} \label{eq:PDE}
  \frac{\partial \bu}{\partial t} + (\bu \cdot \nabla) \bu &=  - \nabla p + \frac{1}{\rm Re}
   \Delta \bu + \bff, \qquad \nabla \cdot \bu = 0, 
\end{align} 
where $\bu$, $p$, and $\bff$ are the velocity, pressure, and forcing term, respectively. Appropriate boundary and initial conditions are needed to close the system.

The FOM is constructed by applying the Galerkin projection of \cref{eq:PDE} onto
the spectral element space with $\mathbb{P}_N$--$\mathbb{P}_{N-2}$
velocity-pressure coupling.
The value of $N$ for each test case in our numerical investigation is specified in Section~\ref{section:numerical-results}. For time discretization, we employ the semi-implicit
BDF$k$/EXT$k$ scheme \cite{fischer2017recent}, which uses the $k$th-order backward differencing (BDF$k$) for the time-derivative, $k$th-order extrapolation (EXT$k$) for the advection and forcing terms, and
an implicit treatment of the dissipation terms. Following 
\cite{fischer2017recent}, we set $k=3$ so that the imaginary eigenvalues
associated with the skew-symmetric advection operator lie within the stability
region of the BDF$k$/EXT$k$. The resulting full discretization leads to solving a linear unsteady Stokes system at each
time step. Further details of the FOM derivation are provided in \cite{tsai2022parametric}. 

\subsection{Galerkin Reduced Order Model (G-ROM)}
    \label{section:g-rom}
    
In this section, we introduce the Galerkin reduced order model (G-ROM).
The reduced basis functions are constructed using the standard proper orthogonal decomposition (POD) procedure
\cite{berkooz1993proper,volkwein2013proper}. Specifically, we collect a set of FOM solutions lifted by the zeroth mode $\bphi_0$,
and assemble their Gramian matrix with respect to the $L^2$ inner product (see,
e.g.,~\cite{fick2018stabilized,kaneko2020towards} for alternative strategies), where the zeroth mode is set to be the time-averaged velocity field over the snapshot interval. 
The first $r$ POD basis functions $\{\bphi_i\}^r_{i=1}$ are constructed from the first $r$
eigenmodes of the Gramian and $\bX^r := \text{span} \{\bphi_i\}^r_{i=1}$ is the ROM space. The G-ROM is then derived by substituting the ROM expansion
\begin{equation} \label{eq:romu}
   \bu_\tr(\bx, t) = \bphi_0(\bx) + \sum_{j=1}^r \romcoef{j}(t) \bphi_j(\bx)
\end{equation}
into the weak form of the NSE \cref{eq:PDE}:
{\em Find $\bu_\tr$ 
such that, for all $\bv \in \bX^r$,}
\begin{eqnarray}
    && 
    \left(
        \frac{\partial \bu_\tr}{\partial t} , \bv_{i} 
    \right)
    + Re^{-1} \, 
    \left( 
        \nabla \bu_\tr , 
        \nabla \bv_{i} 
    \right)
    + \biggl( 
        (\bu_\tr \cdot \nabla) \bu_\tr ,
        \bv_{i} 
    \biggr)
    = 0,  
    \label{eq:gromu}
\end{eqnarray}
where $(\cdot,\cdot)$ denotes the $L^2$ inner product. 

\begin{remark}
    We note that, in the case of fixed geometries, the divergence and pressure
    terms drop out of \eqref{eq:gromu} because the ROM basis functions 
    are weakly divergence-free. For ROMs that include the pressure approximation, see, e.g.,~\cite{ballarin2015supremizer,
    hesthaven2015certified}.
\end{remark}

Let $A$, $B$, and $C$ represent the stiffness, mass, and advection operators,
respectively, with entries
\begin{align} 
    A_{ij} =\int_{\Omega}\nabla \bphi_i :\nabla \bphi_j\, dV, \quad  B_{ij} =\int_{\Omega}\bphi_i\cdot\bphi_j\, dV, \quad 
    C_{ikj}=\int_{\Omega}\bphi_i\cdot(\bphi_k\cdot\nabla)\bphi_j\,dV. \label{eq:Cu} 
\end{align}
With (\ref{eq:gromu}), the ODEs for the ROM coefficients $\romcoef{j}$ are derived: For each $i=1,\ldots,r$,
\begin{align}
	\sum^r_{j=1}B_{ij}\frac{d \romcoef{j}(t)}{dt} & =
	-\sum^r_{k=0}\sum^r_{j=0} C_{ikj} \romcoef{k}(t) \romcoef{j}(t) - {\rm Re^{-1}}
	\sum^r_{j=0} A_{ij} \romcoef{j}(t). \label{eq:nse_ode}
\end{align}

\section{Filter-Based ROM Stabilizations}
    \label{sec:les-rom}

To illustrate the construction of the new StabOp, we use one specific filter-based ROM stabilization, the Leray ROM (L-ROM).
We emphasize, however, that the StabOp idea can be applied to other types of filter-based ROM stabilizations.  
First, in Section~\ref{section:rom-filters}, we  outline the ROM spatial filters generally used to construct filter-based ROM stabilizations. Then, in Section~\ref{section:l-rom}, we outline the L-ROM.

\subsection{ROM Spatial Filters}
    \label{section:rom-filters}

There are three ROM spatial filters in current use: 
(i) the ROM projection \cite{kaneko2020towards,wells2017evolve}; 
(ii) the ROM differential filter (DF) \cite{tsai2022parametric,tsai2025time}; and (iii) the ROM higher-order algebraic filter \cite{tsai2025time}. 
In this paper, we use the first two ROM spatial filters.

\subsubsection{ROM Projection}

The {\it ROM projection} reads: 
Given a positive integer $0 < r_1 < r$ and 
{a velocity field} $\bu_\tr(\bx) = \sum^r_{j=1} \romcoef{j} \bphi_j(\bx)$, find {the filtered velocity field} $\obu_\tr(\bx) = \sum_{j=1}^{r_1} \filtercoef{j} \bphi_j(\bx) \in \bX^{r_1}$
such that 
\begin{eqnarray} 
    \biggl( 
        \obu_\tr , \bphi_i 
        \biggr) 
        = \biggl(\bu_\tr, \bphi_i \biggr)
    \qquad \forall \, i=1, \ldots r_1.
    \label{eq:romproject-weak}
\end{eqnarray}
The ROM projection weak form (\ref{eq:romproject-weak}) is equivalent to setting the last $r-r_1$ ROM coefficients of the input, $\bu_\tr$, to be zero, that is, $\romcoef{j} = 0$ for $j=r_1+1,\ldots,r$.

\subsubsection{ROM Differential Filter}
The {\it ROM differential filter (DF)} reads:
Given 
a velocity field $\bu_\tr(\bx) = \sum^r_{j=1} \romcoef{j} \bphi_j(\bx)$, find the filtered velocity field $\obu_\tr(\bx) = \sum_{j=1}^{r} \filtercoef{j} \bphi_j(\bx)$ such that 
\begin{eqnarray} 
    \biggl( 
        \obu_\tr - \delta^2 \Delta 
        \obu_\tr , \bphi_i 
        \biggr) 
        = \biggl(\bu_\tr, \bphi_i \biggr)
    \qquad \forall \, i=1, \ldots r,
    \label{equation:df-weak}
\end{eqnarray}
where $\delta$ is the filter radius. 
We note that, in contrast with \eqref{eq:romu}, the expansions for $\bu_\tr$ and $\obu_\tr$ do not include the zeroth mode, $\bphi_{0}$. 
The reason for not including $\bphi_{0}$ in our expansions is that this strategy
was shown in~\cite{wells2017evolve} to yield more accurate results. 

The DF weak form~\eqref{equation:df-weak}
yields the following linear system:
\begin{eqnarray} 
    \left( \mathbf{I} + \delta^2 A \right) \filtervec = \romvec,
    \label{equation:df-linear-system}
\end{eqnarray}
where $\filtervec$ and $\romvec$ denote the vectors of ROM coefficients associated with $\obu_\tr$ and $\bu_\tr$, respectively, $\mathbf{I}$ is the identity matrix, and $A$ is the ROM stiffness matrix in \eqref{eq:Cu}  \footnote{In general, the inverse of the ROM mass matrix $B^{-1}$ appears in the DF formulation. However, because the POD basis functions are orthonormal with respect to the $L^2$ inner product, $B$ is the identity matrix and is omitted here for clarity.}.
We emphasize that~\eqref{equation:df-linear-system} is a low-dimensional, $r \times r$ linear system, whose computational overhead is negligible.  Thus, DF will be used in Section~\ref{section:l-rom} to construct the Leray ROM, which increases the ROM accuracy without significantly increasing the computational cost.  

\subsection{Leray ROM (L-ROM)} \label{section:l-rom}

The {\it Leray ROM (L-ROM)}~\cite{kaneko2020towards,wells2017evolve} is inspired from the Leray model, which was first introduced by Jean Leray in 1934 as a theoretical tool to prove the existence of weak solutions of the NSE~\cite{leray1934sur}.
In classical CFD, Leray regularization was first used in~\cite{geurts2003regularization} as FOM stabilization for under-resolved simulations of turbulent flows~\cite{layton2012approximate}.  
As noted in~\cite{guermond2004mathematical,guermond2011entropy},
when a differential filter is used, the Leray model is similar to the NS-$\alpha$ model~\cite{FHT2001}.

The L-ROM modifies the standard G-ROM weak formulation~\eqref{eq:gromu} as follows: Find
$\bu_{\tr}$ of the form~\eqref{eq:romu} such that, $\forall \, i=1, \ldots r,$ 
\begin{eqnarray}
    && 
    \left(
        \frac{\partial \bu_\tr}{\partial t} , \bphi_{i} 
    \right)
    + Re^{-1} \, 
    \left( 
        \nabla \bu_\tr , 
        \nabla \bphi_{i} 
    \right)
    + \biggl( 
        (\obu_\tr \cdot \nabla) \bu_\tr ,
        \bphi_{i} 
    \biggr)
    = 0,   
    \label{eq:l-rom}
\end{eqnarray}
where $\obu_\tr$ is the filtered ROM velocity using the DF~\eqref{equation:df-linear-system}. 
We note, however, that the other ROM spatial filters discussed in Section~\ref{section:rom-filters} could also be used to construct the L-ROM. From (\ref{eq:l-rom}), 
a system of ODEs for the ROM coefficients $\romcoef{j}$ is derived: For each $i=1,\ldots,r$,
\begin{align}
    \sum^r_{j=1}B_{ij}\frac{d \romcoef{j}(t)}{dt}  = -\sum^r_{k=0}\sum^r_{j=0} C_{ikj} \filtercoef{k}(t) \romcoef{j}(t) - {\rm Re^{-1}} \sum^r_{j=0} A_{ij} \romcoef{j}(t). 
    \label{eq:lrom_ode}
\end{align}
We refer to the ODEs in (\ref{eq:lrom_ode}) as L-ROM. 
We also note that the L-ROM is almost identical to the G-ROM (\ref{eq:nse_ode}) except that the ROM coefficients of the advecting field are being filtered.

Leray regularization for ROMs was first used in~\cite{sabetghadam2012alpha} 
for the Kuramoto-Sivashinsky equations.  
For fluid flows, L-ROM was first used in~\cite{wells2017evolve} for the 3D flow
past a circular cylinder at $Re={1000}$.  
Since then, L-ROM has been successfully used as a stabilization technique for various under-resolved flows: the NSE~\cite{girfoglio2021pod,girfoglio2023hybrid}, the stochastic NSE~\cite{gunzburger2019evolve,gunzburger2020leray},  the quasigeostrophic equations~\cite{girfoglio2023linear,girfoglio2023novel}, and the turbulent channel flow~\cite{tsai2025time}.

\section{Data-Driven Stabilization Operator (StabOp)}
    \label{section:d2-les-rom}

Despite the success of ROM stabilizations such as those presented in Section \ref{sec:les-rom},  spatial filters still pose major challenges. The choice of the filter radius (for the DF) and of the reduced dimension (for the ROM projection) is critical and 
significantly influences the accuracy of the results. 
Thus, a natural question is whether we can improve the filter-based ROM stabilizations.
In this section, we propose a novel strategy for the construction of ROM stabilizations, which is based on data-driven modeling.
To this end, we turn the problem on its head, and ask the following question:
\vspace{0.1cm}

\begin{mdframed}
    \textbf{(Q)} \textit{Do we really need ROM spatial filters in order to get accurate ROM stabilizations?} 
\end{mdframed}
\vspace{0.1cm}

\noindent We propose the following answer, which is fundamentally different from the general approaches used to build ROM stabilizations: 
\vspace{0.1cm}

\begin{mdframed}
\textbf{(A1)} \textit{We replace the traditional ROM spatial filters with a novel, data-driven ROM stabilization operator (StabOp) that yields the most accurate ROM.}
\end{mdframed}

The motivation for the answer (A1) is that, after all, we are not really interested in filtering (smoothing out) the flow variables.
Instead, {\it the only goal in developing filter-based ROM stabilizations is to construct ROMs that yield more accurate solutions than the standard G-ROM}. Thus, we can replace the standard ROM filter (used to build classical ROM stabilizations) with any other ROM operator as long as this ROM operator yields more accurate ROMs than the given ROM stabilization. 
In a nutshell, this is the idea underlying the development of the new StabOp and the corresponding StabOp-ROM.

To construct the new StabOp, 
we first need to pose the problem in the appropriate spaces.
To this end, we regard the 
StabOp as a mapping from the ROM space, $\bX^r := \text{span} \{ \bphi_{1}, \ldots, \bphi_{r} \}$, to itself.
Specifically, 
StabOp takes a generic input $\bu_r \in \bX^r$ and maps it to the output $\obu_r \in \bX^r$: 
\begin{eqnarray}
    \bu_r
    := \sum_{j=1}^{r} \romcoef{j} \bphi_j
    \ \xmapsto{\text{StabOp}} \ 
    \obu_r
    := \sum_{j=1}^{r} \filtercoef{j} \bphi_j .
    \label{eqn:reg-rom-1}
\end{eqnarray}
With the notation in \eqref{eqn:reg-rom-1}, answer (A1) can be rephrased as follows:
\vspace{0.1cm}

\begin{mdframed}
\textbf{(A2)} \textit{Given a target quantity of interest (QoI) and a ROM stabilization (e.g., L-ROM), identify the StabOp mapping \eqref{eqn:reg-rom-1} that yields the corresponding ROM stabilization (e.g., StabOp-L-ROM) with the most accurate QoI approximation.}
\end{mdframed}
\vspace{0.1cm}
That is, the goal is to find the optimal (with respect to the given QoI and ROM stabilization) StabOp function $\mF(\cdot\,;\utheta): \real^r \longrightarrow \real^r$, which is characterized by the parameter vector $\utheta$, satisfies 
\begin{equation}
    \mF(\romcoef{j}; \utheta)= \filtercoef{j}, \quad \forall \, j = 1,
    \ldots, r, 
    \label{eqn:d2F}
\end{equation}
and is chosen to minimize the loss function $\mL(\utheta)$ defined as 
\begin{equation}
    \mL(\utheta) := \sum^{\Ntrain}_{i=1}\|\qoi_{\fom}(t_i) -
    \qoi_{\text{StabOp-ROM}}(t_i;\romvec)\|^2 
    + \alpha \|\utheta\|^2. 
\end{equation}
Here $\Ntrain$ is the number of training time instances, and $\qoi_{\text{StabOp-ROM}}(t_i; \ua)$ is the QoI predicted by the StabOp-ROM at time $t_i$,  which depends implicitly on $\utheta$ through the vector of ROM coefficients $\ua$. The regularization term $\alpha \|\utheta\|^2$ is included to mitigate overfitting, and $\alpha>0$ is a user-defined parameter.

A variety of model forms for the novel StabOp, $\mF$, can be considered; see, e.g., operator inference~\cite{kramer2024learning,peherstorfer2016data}, SINDy~\cite{brunton2016discovering,messenger2021weak}, or symbolic regression~\cite{kronberger2024symbolic}. In this work, we explore three representative cases: a linear form, a quadratic form, and a general nonlinear form realized using a neural network (NN). Each model is parameterized by a vector $\utheta$ of trainable weights: 

\begin{align}
    \text{linear form}
    ~ & \mF(\romvec;\utheta)
    \coloneqq \tA \romvec 
    + \uwb, ~\text{where}~
       \utheta = [\tA_{11}~\tA_{12}~\cdots~\tA_{rr}~\tb_1~\tb_2~\ldots~\tb_r]^T
    \in \real^{r^2+r}, 
    \label{eqn:model-form-linear}
    \\[0.2cm]
    \text{quadratic form}
    ~ & \mF(\romvec;\utheta)
    \coloneqq \romvec^T \tB \romvec
    + \tA \romvec
    + \uwb,  \label{eqn:model-form-quadratic}
    \\ & \text{where}~ \utheta 
        = [\tB_{111}~\tB_{112}~\cdots~\tB_{rrr}~\tA_{11}~\tA_{12}~\cdots~\tA_{rr}~\tb_1~\tb_2~\cdots~\tb_r]^T \in \real^{r^3+r^2+r}, \nonumber
    \\[0.2cm]
    \text{nonlinear form} 
    ~ & \mF(\romvec;\utheta) = W^{(n_L)} L^{(n_L-1)} + b^{(n_L)}, \label{eqn:model-form-neural-network}  
    \\ & \text{where}~ 
    L^{(\ell)} = h\!\left(W^{(\ell)} L^{(\ell-1)} + b^{(\ell)}\right), 
    \quad \ell = 1, \dots, n_L-1,~\text{and}~
    L^{(0)} = \romvec, \nonumber \\ & \utheta = \text{vec}\left(\{W^l,b^l\}^{n_L}_{l=1}\right) \in \real^{N_{\theta}} 
    \nonumber
    ~\text{denotes the vector of network parameters} \nonumber \\ & \text{obtained by vectorizing all weights and biases $\{W^l,b^l\}^{n_L}_{l=1}$.}
    \nonumber
\end{align}
In \eqref{eqn:model-form-neural-network}, $n_L$ is the total number of layers, including the output layer, ${L}^l$ is the output of the $l$-th layer, $h(\cdot)$ is the activation function, $\{{W}^l,b^l\}^{n_L}_{l=1}$ are the trainable weights and biases, and $N_{\theta}$ is the total number of trainable parameters.

To determine the new StabOp $\mF$ in \eqref{eqn:d2F}, that is, to determine the parameters $\utheta$ in the linear, quadratic, or nonlinear model forms~\eqref{eqn:model-form-linear}--\eqref{eqn:model-form-neural-network},
we solve the following {\it PDE-constrained optimization} problem: 
\begin{equation}
\begin{aligned}
    & \min_{\utheta} \mL(\utheta) \\ 
    & {\text{subject to }} \romvec~\text{solving StabOp-ROM}.
    \label{eqn:reg-rom-energy}
\end{aligned}
\end{equation}

{
We note that linear, quadratic, and nonlinear model forms have been successfully used in ROM closure modeling \cite{mou2021data}, as well as in other data-driven ROMs \cite{barnett2022neural,geelen2023operator, de2026nonlinear}. 
However, our current approach differs fundamentally from these works, as we formulate and solve a PDE-constrained optimization problem to determine the model operators. 
This distinguishes our method from the strategies used in \cite{barnett2022neural,de2026nonlinear,geelen2023operator,mou2021data}, which do not involve PDE-constrained formulations. 
Nonetheless, the success of these earlier works demonstrates that relatively simple parameterizations, such as linear and quadratic forms, can be effective in capturing complex dynamics, particularly in ROM closure modeling and stabilization. 

Although the optimization problem is a PDE-constrained
optimization problem, its solution is carried out entirely at the
ROM level. As a result, the computational cost of the optimization does not scale
with the number of degrees of freedom of the FOM, but only
with the ROM dimension and the chosen parameterization of the
StabOp, $\mF$. Efficient algorithms for PDE-constrained optimization include adjoint-based methods \cite{kim2023generalizable,sirignano2023pde}, Gauss--Newton approaches with forward sensitivity method \cite{ahmed2023forward}, and automatic-differentiation-based techniques. In this work, we consider an automatic-differentiation-based approach. Specifically, we employ reverse-mode automatic differentiation, as
implemented in modern machine-learning frameworks such as
\texttt{PyTorch}, \texttt{JAX}, and \texttt{TensorFlow}, to compute gradients of the loss function with respect to the model parameters
without explicitly deriving adjoint or sensitivity equations.
This choice provides a straightforward and flexible mechanism for
constructing the StabOp-ROM. While convenient and general, reverse-mode automatic differentiation
requires storing the full computational graph of the forward simulation,
which can lead to increased memory usage for problems with long time horizons or when $\mF$ is parameterized by deep neural networks.
In such settings, adjoint-based optimization methods are particularly well suited, as they avoid storing the full forward trajectory and yield gradient computations whose cost is independent of the parameter dimension.

We also note that the ROM differential filter \eqref{equation:df-weak} is a special case of the linear ROM operator \eqref{eqn:model-form-linear}. 
In contrast, the quadratic ROM operator \eqref{eqn:model-form-quadratic} and the neural network ROM operator \eqref{eqn:model-form-neural-network} are nonlinear and, thus, more general than the ROM differential filter.} 

Finally, we note that both the new StabOp $\mF$~\eqref{eqn:d2F} 
and the classical ROM spatial filters (e.g., the ROM differential filter) have the same goal:
Both ROM operators aim at improving the accuracy of the standard G-ROM. 
We emphasize, however, that the two ROM operators use fundamentally different strategies to attain this goal:
Indeed, the classical ROM differential filter is a spatial filter that smooths out the small spatial structures in the input.
In contrast, the new StabOp $\mF$ is a generic ROM operator that, a priori, is not known to be a spatial filter (see Remark~\ref{remark:d2F-filter}). 
Indeed, $\mF$ could in principle even increase certain components of the input as long as the StabOp-ROM built by using $\mF$ yields accurate results (see Section~\ref{section:numerical-results-spatial-filter} for an example).
A numerical investigation of whether the new StabOp $\mF$ is a spatial filter is performed in Section~\ref{section:numerical-results-spatial-filter}.

\subsection{Data-Driven 
Stabilization Operator for
Leray ROM (StabOp-L-ROM)}
\label{subsection:d2lrom}

In this section, we introduce a new data-driven Leray-ROM by applying the data-driven modeling strategy developed above to the classical L-ROM described in Section~\ref{section:l-rom}. 
To this end, we first replace the classical L-ROM~\eqref{eq:l-rom} with a more general form, which we denote {\it StabOp-L-ROM}, in which the traditional ROM spatial filter is replaced with the new 
StabOp, $\mF$, which has one of the model forms in (\ref{eqn:model-form-linear})--(\ref{eqn:model-form-neural-network}): 
\begin{align}
    \sum^r_{j=1}B_{ij}\frac{d \romcoef{j}(t)}{dt}  = -\sum^r_{k=0}\sum^r_{j=0} C_{ikj} \mF(\romvec(t);\utheta)_{k} \romcoef{j}(t) - {\rm Re^{-1}} \sum^r_{j=0} A_{ij} \romcoef{j}(t). 
    \label{eq:d2-leray-rom}
\end{align}

We note that the classical L-ROM~\eqref{eq:l-rom} 
is a special case of the more general L-ROM form~\eqref{eq:d2-leray-rom}. 
Indeed, as noted in Section~\ref{section:rom-filters}, the ROM differential filter~\eqref{equation:df-weak} corresponds to a special case of the linear operator form~ \eqref{eqn:model-form-linear}. Since the new StabOp used to construct \eqref{eq:d2-leray-rom} can be nonlinear (see \eqref{eqn:model-form-quadratic} and \eqref{eqn:model-form-neural-network}), the new L-ROM is more general than the classical L-ROM. 

To construct the new StabOp-L-ROM (\ref{eq:d2-leray-rom}), we need to find the vector of parameters $\utheta$ that determines the particular form of the StabOp, $\mF$. 
To this end, we solve the following PDE-constrained optimization problem: 
\begin{equation}
\begin{aligned}
    & \min_{\utheta} \ \sum_{i=1}^{\Ntrain} \| \qoi_{\fom}(t_i) - \qoi_{
    \text{StabOp-L-ROM}}(t_i;\romvec) \|^2 
    + \alpha\|\utheta\|^2 
    \\ & \text{subject to } \romvec~
    \text{solving StabOp-L-ROM (\ref{eq:d2-leray-rom}).} 
    \label{eqn:StabOp-L-ROM-2}
\end{aligned}
\end{equation}
To summarize, the new StabOp-L-ROM 
consists of the system of ODEs \eqref{eq:d2-leray-rom} in which StabOp has one of the model forms in \eqref{eqn:model-form-linear}--\eqref{eqn:model-form-neural-network} and is equipped 
with the optimal parameters $\utheta$ found in \eqref{eqn:StabOp-L-ROM-2}.

\begin{remark}[Is the new StabOp a filter?]
\label{remark:d2F-filter}
By construction, the novel StabOp, $\mF$, defined in~\eqref{eq:d2-leray-rom}--\eqref{eqn:StabOp-L-ROM-2} yields an optimally accurate L-ROM stabilization. 
We could, however, ask the following natural question: 
Does $\mF$ represent a ROM spatial filter?
In other words, in \eqref{eqn:reg-rom-1}, is $\obu_r$ smoother than $\bu_r$?
And if so, how does this new 
data-driven ROM spatial filter, $\cF$, compare with other ROM spatial filters, e.g., the ROM projection or the ROM differential filter?
In our numerical investigation in Section~\ref{section:numerical-results-spatial-filter}, we address these important questions. 
\end{remark}

\section{StabOp Implementation} 
\label{section:numerical-method}

As outlined in \textbf{(A2)} (Section \ref{section:d2-les-rom}), we develop a data-driven ROM stabilization, StabOp-L-ROM, which provides an accurate approximation of our \qoi{}. To achieve this goal, we should carefully select the StabOp, $\mF(\ua;\utheta)$. 
As pointed out in Section \ref{section:d2-les-rom}, various choices are possible, with the simplest being the linear and quadratic models in \eqref{eqn:model-form-linear} and \eqref{eqn:model-form-quadratic}, respectively. However, in high-Reynolds-number test cases, these may fail to accurately capture 
the complex dynamics, especially in the extrapolation regime.
In such cases, we rely on nonlinear operators, such as neural networks, as introduced in \eqref{eqn:model-form-neural-network}.

The algorithm used to solve the PDE-constrained optimization problem~\eqref{eqn:StabOp-L-ROM-2} for constructing the StabOp, $\mF(\ua;\utheta)$, is summarized in Algorithm~\ref{alg:d2lrom-training}. 
The training procedure involves the selection of two hyperparameters:
the differential filter radius $\delta$ and the $L^2$ regularization
weight $\alpha$.
The parameter $\delta$ controls the amount of filtering applied through
the ROM differential filter~\eqref{equation:df-linear-system}, which is used 
to initialize the operator parameters (see Remark~\ref{remark-training} for more details), while $\alpha$ penalizes large
values of $\utheta$ in the loss function and thereby mitigates overfitting.
We denote by $\mP_\delta$ and $\mP_\alpha$ the discrete search spaces for
$\delta$ and $\alpha$, respectively.

\begin{algorithm}[!ht]
    \caption{Training of StabOp, $\mF(\ua;\utheta)$}
    \label{alg:d2lrom-training}
    \begin{algorithmic}[1]
    \State \textbf{Input:} Initial condition $\ua(0)$, ground truth $\qoi_{\fom}(t)$ trajectory, initial step size parameter $\eta$, number of epochs $\Nepoch$, hyperparameter spaces for filter radius $\mP_\delta$ and $L^2$ regularization weight $\mP_{\alpha}$
    \For{$(\delta, \alpha) \in \mP_{\delta} \times \mP_{\alpha}$} 
    \State Initialize network parameters $\utheta_{(0)}$ using 
    the ROM differential filter ~\eqref{equation:df-linear-system} with parameter $\delta$
    \For{$n=1,\dots, \Nepoch$} 
        \State Solve the StabOp-L-ROM (\ref{eq:d2-leray-rom}) with $\mF(\cdot\,;\utheta_{(n-1)})$ and evaluate and store $\qoi_{\text{StabOp-L-ROM}}$ over the training interval $[\tinit, \ttrain]$ 
    \State Compute the loss:
        $\mL(\utheta) = \sum_{i=1}^{\Ntrain} \left\| \qoi_{\text{StabOp-L-ROM}}(t_i) - \qoi_{\fom}(t_i) \right\|^2 + \alpha\|\utheta\|^2
       $
    \State Compute the gradient $\nabla_{\utheta} \mL(\utheta)$
    \State Update parameters using an L-BFGS step: $\utheta_{(n)} \gets \mathrm{LBFGS}(\utheta_{(n-1)};\mL,\nabla\mL)$

    \State Solve the StabOp-L-ROM with $\mF(\cdot\,;\utheta_{(n)})$ and evaluate and store $\qoi_{\text{StabOp-L-ROM}}$ over the validation interval $[\tinit, \tval]$ 
    
    \State Evaluate the validation loss $\mathcal{E}_{\mathrm{val}}$ (\ref{eq:adjusted-r2})

    \If{$\mathcal{E}_{\mathrm{val}}$ is smaller than the previous one}
    \State Update the best network parameter $\utheta_{\text{best}}$ with $\utheta_{(n)}$
    \EndIf

    \EndFor
    \State Solve the StabOp-L-ROM with $\mF(\cdot\,;\utheta_{\text{best}})$ and evaluate and store $\qoi_{
    \text{StabOp-L-ROM}}$ over the target interval $[\tinit, \tfinal]$ 
    \If{the predicted $\qoi_{
    \text{StabOp-L-ROM}}$ satisfies the growth constraint (\ref{eq:bounded_growth})}
    \State Evaluate the validation loss $\mathcal{E}_{\mathrm{val}}$ (\ref{eq:adjusted-r2})
    \If{$\mathcal{E}_{\mathrm{val}}$ is less than the the optimal $\mathcal{E}_{\mathrm{val}}$}
    \State Update the final network parameter $\utheta$ with $\utheta_{\text{best}}$
    \EndIf
    \EndIf
\EndFor
\State \textbf{Output:} Trained parameters $\utheta$ that define 
the StabOp, $\mF$ 
\end{algorithmic}
\end{algorithm}

In Algorithm~\ref{alg:d2lrom-training}, for each candidate pair $(\delta,\alpha)\in\mP_\delta\times\mP_\alpha$, the loss function $\mL(\utheta)$ is minimized using the L-BFGS optimizer implemented in PyTorch, which iteratively updates the model parameters $\utheta$ using gradients computed via reverse-mode automatic differentiation and a line-search procedure. Hyperparameter selection is based jointly on a validation loss and a growth constraint.

The validation loss is defined to balance accuracy and variability. It is known that the following widely used validation metric
\begin{equation}
     R^2_{\mathrm{val}} = 1 - \frac{\sum_{k=1}^{\Nval} \bigl(\qoi_{\fom}(t^k) - \qoi_{\rom}(t^k) \bigr)^2}
    {\sum_{k=1}^{\Nval} \bigl(\qoi_{\fom}(t^k) - \overline{\qoi}_{\mathrm{val}} \bigr)^2},
    \label{eqn:R2}
\end{equation}
with $\overline{\mathrm{QoI}}_{\mathrm{val}}$ denoting the mean of $\qoi_{\fom}$ over $\Tval = [\ttrain,\tval]$,
can yield misleadingly high values when the validation data exhibit low variability. In such cases, the denominator of $R^2_{\mathrm{val}}$ becomes small, and a dissipative model that stays close to the mean trajectory can achieve a deceptively high score despite underpredicting the temporal fluctuations of the validation data. Similar limitations in time-series model evaluation have been reported in \cite{onyutha2020r,onyutha2022hydrological}, where it was shown that $R^2_{\mathrm{val}}$ can yield spuriously high values even when the discrepancies between model and observations are large, particularly in situations with low variance or strong autocorrelation. 
Motivated by these findings and by our own preliminary results using $R^2_{val}$ as the validation metric, we augment it with the error in the variance
\begin{equation}
    \epsilon_{\mathrm{std}, \mathrm{val}} = \frac{\bigl|\sigma_{\mathrm{val}}(\qoi_{\rom}) -
    \sigma_{\mathrm{val}}(\qoi_{\fom})\bigr|}{\sigma_{\mathrm{val}}(\qoi_{\fom})},
\end{equation}
where $\sigma_{\mathrm{val}}(\cdot)$ denotes the standard deviation of the corresponding 
samples restricted $\Tval$.
This yields the following combined validation loss
\begin{equation}
\label{eq:adjusted-r2}
    \valmetric = -R^2_{\mathrm{val}}
    + \mu \epsilon_{\mathrm{std},\mathrm{val}}, 
\end{equation}
where parameter $\mu$ balances accuracy and variability: smaller values of $\mu$ favor a high $R^2_{\mathrm{val}}$, while larger values of $\mu$ prioritize matching the FOM variability, thereby preventing false selections that would otherwise favor overly dissipative models. 
In all the numerical tests in Section~\ref{section:numerical-results}, we set $\mu=2.5$, which is the value that produced the best results in our numerical investigation. 

The growth constraint is defined as 
\begin{equation}
\frac{\max_{t \in [t_{\mathrm{init}},\,t_{\mathrm{final}}]}
  \bigl|\qoi_{\rom}(t) - \overline{\qoi}_{\mathrm{train}} \bigr|}{\max_{1 \leq k \leq N_{\mathrm{train}}} 
  \bigl| \qoi_{\fom}(t^k) - \overline{\qoi}_{\mathrm{train}} \bigr|,} < \egrowth, \label{eq:bounded_growth}
\end{equation}
where $\overline{\qoi}_{\mathrm{train}}$ is the mean of the $\qoi_{\fom}$ over $\Ttrain$. 
This condition requires that the maximum deviation of the StabOp-L-ROM predicted QoI from $\overline{\qoi}_{\mathrm{train}}$ over the entire interval $[\tinit,\tfinal]$ remains bounded 
by the largest deviation observed in the FOM QoI in the training interval, up to a tolerance $\egrowth$.
In all the numerical experiments in Section \ref{section:numerical-results}, we set
$\egrowth=2$.

\begin{remark}
    We note that although StabOp-L-ROM and L-ROM are simulated
    over the target interval $[\tinit, \tfinal]$ during the training, no testing data are used to tune the hyperparameters. The growth
    constraint is employed solely to discard candidate models that match the validation data but become unstable outside the validation region \cite{gahr2024scientific}.
\end{remark}

We also note that the training procedure in Algorithm~\ref{alg:d2lrom-training} does not require using physics-informed neural networks (PINNs)~\cite{raissi2019physics}, as the loss is evaluated on output quantities of interest rather than enforcing the governing equations through the neural network.

In the numerical assessment of the new StabOp-L-ROM in Section~\ref{section:numerical-results}, we compare it with the classical L-ROM~\eqref{eq:l-rom} equipped with the ROM differential filter \eqref{equation:df-weak} and an optimal filter radius, $\delta$.
To ensure a fair comparison between the new StabOp-L-ROM and the classical L-ROM, we cast the L-ROM as a learning process that uses the same validation loss $\mathcal{E}_{\mathrm{val}}$ (\ref{eq:adjusted-r2}) and growth bound (\ref{eq:bounded_growth}) as those used in the training of StabOp-L-ROM. Algorithm~\ref{alg:lrom-training} summarizes the hyperparameter selection procedure for the standard L-ROM, outlined in Section~\ref{section:l-rom}. For each candidate filter radius $\delta \in \mP_\delta$, the L-ROM is simulated over the target interval $[\tinit,\tfinal]$. If the predicted trajectory satisfies the bounded growth condition, the validation loss $\valmetric$ is computed. The filter radius $\delta$ that minimizes $\valmetric$ among the admissible candidates is selected as the optimal filter radius. 
\begin{algorithm}
    \caption{Finding optimal filter radius, $\delta$, for L-ROM} \label{alg:lrom-training}
    \begin{algorithmic}[1]
    \State \textbf{Input:} Initial condition $\ua(0)$, ground truth $\qoi_{\fom}(t)$ trajectory, hyperparameter space for filter radius $\mP_\delta$
    \For{$\delta \in \mP_{\delta}$} 
    \State Solve L-ROM (\ref{eq:l-rom}) with the ROM differential filter and filter radius $\delta$, and evaluate and store $\qoi_{
    \text{L-ROM}}$ over the target interval $[\tinit, \tfinal]$ 

    \If{the predicted $\qoi_{
    \text{L-ROM}}$ satisfies the growth constraint (\ref{eq:bounded_growth})}
    \State Evaluate the validation loss $\mathcal{E}_{\mathrm{val}}$ (\ref{eq:adjusted-r2})
    \If{$\mathcal{E}_{\mathrm{val}}$ is less than the the optimal $\mathcal{E}_{\mathrm{val}}$}
    \State Set $\delta_{\text{optimal}} \gets \delta$
    \EndIf
    \EndIf
\EndFor
\State \textbf{Output:} Optimal filter radius $\delta_{\mathrm{optimal}}$ for L-ROM
\end{algorithmic}
\end{algorithm}

\begin{remark}
\label{remark-training}
{In challenging numerical experiments (such as those in Section~\ref{section:numerical-results}), the initialization of StabOp, $\mF$, plays an important role in ensuring stable PDE-constrained optimization. In all test cases, except the 2D flow past a cylinder (which did not require a particular initialization to guarantee stability), we leverage the ROM differential filter associated with a prescribed filter radius $\delta$, $(\mathbb{I}+\delta^2 A)^{-1}$, as an initialization mechanism. For the linear and quadratic model forms, the differential filter can be incorporated in two ways. In the first approach, the trainable linear operator $\tilde{A}$ in \eqref{eqn:model-form-linear} (linear) or \eqref{eqn:model-form-quadratic}} (quadratic) is initialized directly with the differential filter.
In the second approach, the model is augmented by introducing an
additional linear operator initialized as the differential filter and added explicitly to the model form. In both cases, the differential filter is trainable and updated during
optimization. For the nonlinear model, we employ the augmented formulation, 
in which the differential filter appears as an explicit additive component alongside the network output. 
In the nonlinear model results presented in Section~\ref{section:numerical-results}, the
differential filter component is held fixed during training and serves
solely as an initialization mechanism.
While this component could in principle be treated as trainable, we keep
it fixed in this work and leave the investigation of trainable variants
for future study. Overall, this initialization strategy provides a stable starting point for training StabOp, especially in the under-resolved regime. 
\label{remark:initialization}
\end{remark}

\section{Numerical Results}
    \label{section:numerical-results}

In this section, we evaluate the performance of the proposed {\dlrom} (Section~\ref{subsection:d2lrom}) with linear, quadratic, and nonlinear model forms, and compare it with the classical L-ROM (\ref{eq:l-rom}) equipped with an optimal differential filter~\eqref{equation:df-weak}, and the classical G-ROM. 
For clarity, we will refer to the new StabOp-L-ROMs with linear, quadratic, and nonlinear model forms as \textbf{StabOp-L-ROM (linear)}, \textbf{StabOp-L-ROM (quad)}, and \textbf{StabOp-L-ROM (NN)}, respectively.
The numerical comparison is conducted across four test problems:
(i) 2D flow past a cylinder (Section \ref{results-cyl}),
(ii) 2D lid-driven cavity (Section \ref{results-ldc}),
(iii) 3D flow past a hemisphere (Section \ref{results-hemi}), and 
(iv) 3D minimal channel flow (Section \ref{results-minimal}).
All test cases are set in the under-resolved regime, where the number of ROM basis functions $r$ is insufficient to fully capture the underlying dynamics.
The only exception is the 2D flow past a cylinder test case, which is investigated in the resolved regime.
As discussed in Section~\ref{results-cyl},  this choice is motivated by the fact
that an under-resolved ROM arises for this problem only in the unrealistic case
$r=1$. For the remaining test cases, we evaluate the ROMs for several $r$ values in the under-resolved regime.
For these $r$ values, we also monitor their corresponding energy threshold, $\delta_\sigma \in [0, 1]$, which is given by the following formula: 
\begin{equation}
    \frac{\sum^{r}_{i=1}\lambda_i}{\sum^{\Ntrain}_{i=1}\lambda_i} \ge \delta_\sigma , \label{eq:energy_criteria}
\end{equation}
where $\lambda_i$ is the $i$-th largest eigenvalue of the snapshot Gramian matrix using the $L^2$ inner product, and $\Ntrain$ is the number of snapshots. We emphasize that the under-resolved and convection-dominated regimes are representative of realistic engineering and geophysical settings, in which the efficient and accurate numerical simulation of turbulent flows is critical.

In our numerical investigation, 
the new StabOp-L-ROM is trained by using Algorithm~\ref{alg:d2lrom-training}, with hyperparameters specified separately for each test problem. 
We note that, for the nonlinear model form 
\eqref{eqn:model-form-neural-network}, we use a two-layer feed-forward fully-connected neural network ($n_L=2$) with $r$ neurons in each hidden layer for all test cases. 
Therefore, the resulting number of parameters (i.e., the dimension of the vector $\utheta$) depends only on the chosen ROM dimension, $r$.

The L-ROM is trained by using Algorithm~\ref{alg:lrom-training} with the filter radius selected from a hyperparameter space $\mP_{\delta}$ consisting of $60$ values: twenty values uniformly sampled from each of the intervals $[0.001, 0.01]$, $[0.01, 0.1]$, and $[0.1, 1]$. A large number of values for the filter radius are considered for L-ROM to ensure a thorough search for the optimal 
filter radius.

Despite its relevance to realistic applications in which efficient numerical simulations are critical, the under-resolved regime may seem inadequate to provide accurate flow field approximations. 
Indeed, one could ask whether a relatively low-dimensional (e.g., 10-dimensional) ROM is able to accurately approximate a complex turbulent flow, such as those we consider in this section.
We emphasize that the goal in reduced order modeling of turbulent flows is {\it not} to accurately approximate the {\it pointwise}, fine structures of the flow, such as those displayed by a high-resolution FOM with millions or even billions of degrees of freedom.
Instead, the ROMs aim at accurately approximating appropriate {\it QoIs}, such as the kinetic energy defined in \eqref{eq:ene}--\eqref{eq:ene_rom}, which are {\it integrated} quantities.
For those QoIs, there is hope that a relatively low-dimensional ROM could yield accurate approximations.

In this section, we demonstrate that the new {\dlrom} achieves this goal. In our numerical study, we consider the kinetic energy as the QoI. The FOM kinetic energy is defined as 
\begin{equation}
    \EFOM(t) = \frac{1}{2} \buu^T \mathcal{M}  \buu \approx \frac{1}{2}\int_{\Omega} \bu \cdot \bu ~d\Omega,
    \label{eq:ene} 
\end{equation}
where $\buu$ denotes the spectral element coefficient vector of the velocity field $\bu$, and $\mathcal{M}$ is the FOM mass matrix. The ROM kinetic energy is defined as 
\begin{equation}
    \EROM(t) = \frac{1}{2} \ua^T B \ua 
    \approx 
    \frac{1}{2}\int_{\Omega} 
   \bu_\tr \cdot \bu_\tr ~d\Omega
   , \label{eq:ene_rom}
\end{equation}
where $\ua \in \real^r$ is the ROM state and $B$ the ROM mass matrix in \eqref{eq:Cu}. While this study focuses on kinetic energy, other QoIs such as POD coefficient trajectories, drag, turbulent kinetic energy, Nusselt number, or Reynolds stresses can also be used in Algorithm~\ref{alg:d2lrom-training} to train the new StabOp.  

Moreover, we compare the novel StabOp-L-ROMs with the L-ROM and the G-ROM in terms of the energy spectrum computed from the spatial velocity field at a fixed time.
To this end, the velocity field is interpolated onto a uniform Cartesian grid, and a Hann window is applied in each spatial direction to reduce spectral leakage while preserving energy. 
The Fourier transform of each velocity component is then computed.
Physical wavenumbers $\boldsymbol{k}$ are defined from the grid spacing as $k_{i} = \frac{2\pi}{dx_i}$, for all the components $i=1, \dots, d$, where $d$ is the test case dimensionality ($2$ or $3$ in our test cases).
By doing so, each mode corresponds to a true spatial frequency. 
The spectral energy density can be obtained as $E(\boldsymbol{k}) = \frac{1}{2}(\sum_{i=1}^d|\hat{u}_{i}|^2)$, where $\hat{u}_i$ indicates the Fourier transform of the $i$-th velocity component.
An isotropic energy spectrum is finally constructed by averaging $E(\boldsymbol{k})$ over circular (2D) or spherical (3D) shells of constant wavenumber magnitude 
$\kappa = |\boldsymbol{k}|$.

\subsection{2D Flow Past a Cylinder}
\label{results-cyl}

Our first test problem is the 2D flow past a cylinder at the Reynolds number
$\rm Re=500$. The computational domain is 
$\Omega = [-2.5,~17] \times [-5,~5]$, where the cylinder has
unit diameter and is centered at the origin. 
We focus on the time interval $[500,~600]$, measured in convective time units based on the free-stream velocity, after the von Karman vortex shedding is developed. 

\begin{figure}[!ht]
    \centering
    \includegraphics[width=1\textwidth]{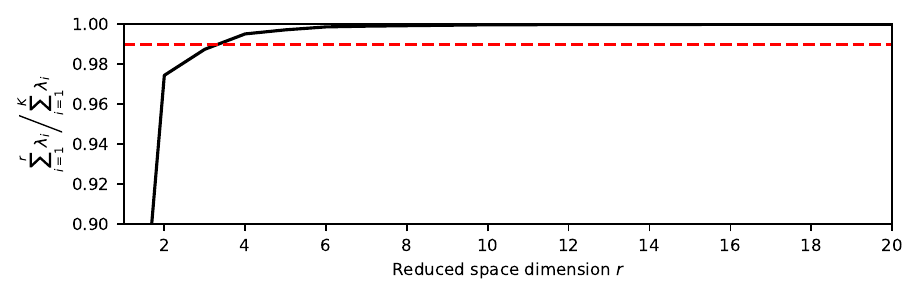}
    
   \caption{2D flow past a cylinder at $\rm Re={500}$. The behavior of $\sum^{r}_{i=1} \lambda_i /\sum^{\Ntrain}_{i=1} \lambda_i$ as a function of the ROM space dimension, $r$.}
   \label{fig:2dcyl_emr}
\end{figure}

The ROM basis functions $\{\bphi_i\}^r_{i=1}$ are constructed via POD by using $\Ntrain=2001$ snapshots from the training interval $\Ttrain = [500,~520]$, which corresponds to a frequency of $0.01$. The zeroth mode $\bphi_0$ is defined as the velocity at $t=500$, and the ROM initial condition is obtained by projecting the
lifted snapshot at $t=500$ onto the reduced space. We consider four reduced space dimensions, $r=4, 6, 8, 10$.
These values correspond to energy thresholds $\delta_\sigma>0.99$ in \eqref{eq:energy_criteria}, as shown in Fig.~\ref{fig:2dcyl_emr}. 
From Fig.~\ref{fig:2dcyl_emr}, we observe that for $r=2$ the reduced space already captures nearly $98\%$ of the snapshot energy. Consequently, only the case $r=1$, which captures $73.62\%$ of the snapshot energy, can be regarded as under-resolved. Since an $r=1$ ROM is of limited practical interest, we do not consider that case. 
Thus, for the flow past a cylinder test case, we consider only the resolved regime. 

The new StabOp, $\mF$, in {\dlrom} is trained using the hyperparameters listed in Table \ref{tab:data-cylinder}. The training loss compares the FOM kinetic energy and the predicted kinetic energy from {\dlrom}: 
\begin{equation}
\sum_{i=1}^{\Ntrain} \| \EFOM(t_i) - 
E_{\dlrom}(t_i) \|^2.
\label{eq:loss_ke}
\end{equation}

\begin{table}[htpb!]
    \centering
    \setlength{\tabcolsep}{8pt}
    \renewcommand{\arraystretch}{1.2}
    \begin{tabular}{>{\centering\arraybackslash}m{0.8cm} 
                    >{\centering\arraybackslash}m{1.2cm} 
                    >{\centering\arraybackslash}m{1.cm} 
                    >{\centering\arraybackslash}m{1.cm}
                    >{\centering\arraybackslash}m{1.cm}
                    >{\centering\arraybackslash}m{1cm}
                    >{\centering\arraybackslash}m{0.8cm}
                    >{\centering\arraybackslash}m{1cm}
                    >{\centering\arraybackslash}m{2.5cm}
                    }
    \toprule
\multirow{2}{*}{$r$}&\multirow{2}{2cm}{Energy retained}&\multicolumn{3}{c}{Time window}&\multirow{2}{*}{$h$ (\texttt{NN})}&\multirow{2}{*}{$\eta$}&\multirow{2}{*}{$\Nepoch$}&\multirow{2}{*}{$\alpha$}\\
\cmidrule{3-5}
&& $\Ttrain$ & $\Tval$ & $\Ttest$ &&&&\\
\midrule
4&0.9952&\multirow{5}{*}{$[500,~520]$}&\multirow{5}{*}{$[520, ~540]$}&\multirow{5}{*}{$[540,~600]$}&\multirow{5}{*}{ReLU}&\multirow{5}{*}{$0.2$}&\multirow{5}{*}{$100$}&\multirow{5}{2cm}{optimized in $\mP_\alpha= \{10^{-2}, \allowbreak 10^{-4}, \allowbreak 10^{-6}, \allowbreak 10^{-8}\}$}\\
\cmidrule{1-2}
6&0.9987&&&&&&&\\
\cmidrule{1-2}
8&0.9993&&&&&&&\\
\cmidrule{1-2}
10&0.9997&&&&&&&\\
    \bottomrule
    \end{tabular}
    \caption{2D flow past a cylinder at $\rm Re={500}$. Hyperparameters for the {\dlrom}s (linear, quadratic, and fully nonlinear). Columns labeled \texttt{NN} correspond to hyperparameters specific to the neural-network model, namely the activation function.}
    \label{tab:data-cylinder}
\end{table}

We note that the same $L^2$ regularization weight, $\alpha$, is used for the linear, quadratic, and nonlinear operators. In addition, in our numerical investigation, we use the model forms presented in \eqref{eqn:model-form-linear}--\eqref{eqn:model-form-neural-network}. A different model form or initialization employing the ROM differential filter (as in Remark~\ref{remark:initialization}) is not needed in this test case since the optimization process is already stable. 
For L-ROM, the ROM differential filter radius hyperparameter space described at the beginning of this section is 
used.

\begin{figure}[!ht]
    \centering
    \includegraphics[width=0.9\textwidth]{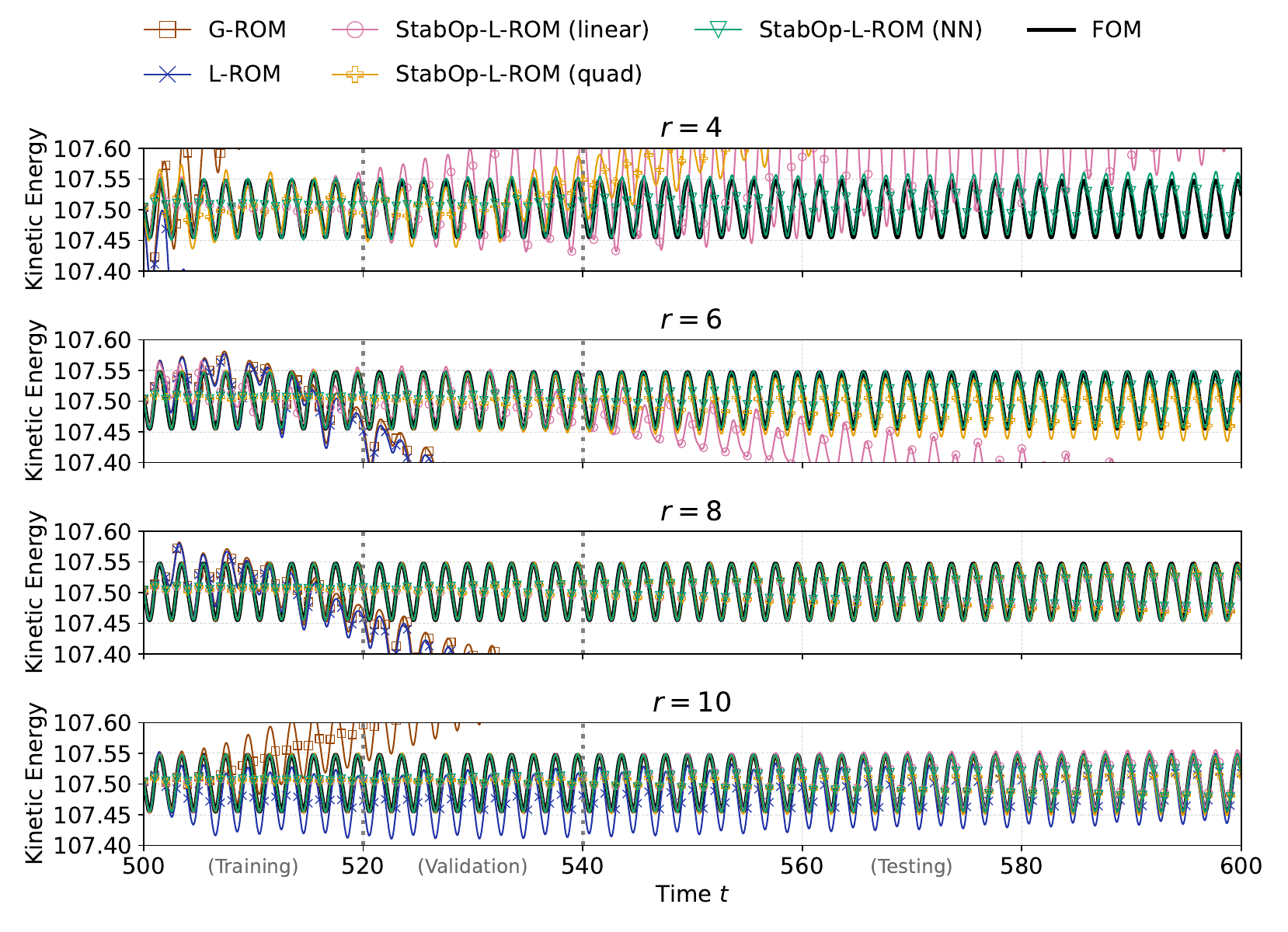}
    \caption{2D flow past a cylinder at $\rm Re=500$. Kinetic energy evolution of the new {\dlrom}, along with the results of the G-ROM, L-ROM with an optimal filter radius, and FOM.} 
    \label{fig:2dfpc_ke}
\end{figure}

Figure~\ref{fig:2dfpc_ke} shows the kinetic energy evolution on the target interval $[500,~600]$ for the four ROM dimensions investigated.  The training, validation, and test intervals are indicated in the figure by vertical dotted lines. 
The comparison includes five ROMs, that is, G-ROM, L-ROM, and {\dlrom} with linear, quadratic, and nonlinear model forms, along with the FOM. 

For $r=4$ and $r=6$, both the G-ROM and the L-ROM with an optimal filter radius are unstable and fail to provide accurate long-time predictions, displaying significant deviations from the FOM. For $r=4$, both {\dlrom} (linear) and {\dlrom} (quadratic) exhibit better stability than L-ROM and G-ROM: the {\dlrom} kinetic energy remains accurate in $\Ttrain$ although it starts to deviate in $\Tval$. In contrast, the {\dlrom} (NN) is the model that remains stable and accurate in both $\Tval$ and $\Ttest$. For $r=6$, both 
{\dlrom} (quadratic) and {\dlrom} (NN) accurately approximate the FOM, while {\dlrom} (linear) starts to deviate in $\Ttest$.
For higher dimensions (e.g., $r=8$ and $r=10$), the G-ROM remains unstable, and the L-ROM improves stability only for $r=10$, where it shows a small deviation from the FOM. In contrast, all three variants of the StabOp-L-ROM yield stable and accurate results over the entire target interval.

Overall, these results demonstrate that the {\dlrom} improves stability and accuracy compared to the L-ROM with an optimal filter radius, and that incorporating a quadratic or nonlinear StabOp yields better accuracy than the linear form, especially for smaller reduced dimensions. 
Moreover, these results indicate that the classical L-ROM cannot improve the G-ROM's accuracy (see the low ROM dimension cases, i.e., $r=4$ and $r=6$), even when an optimal filter radius is used to build the L-ROM. 

\begin{figure}[!ht]
    \centering
    \includegraphics[width=\linewidth]{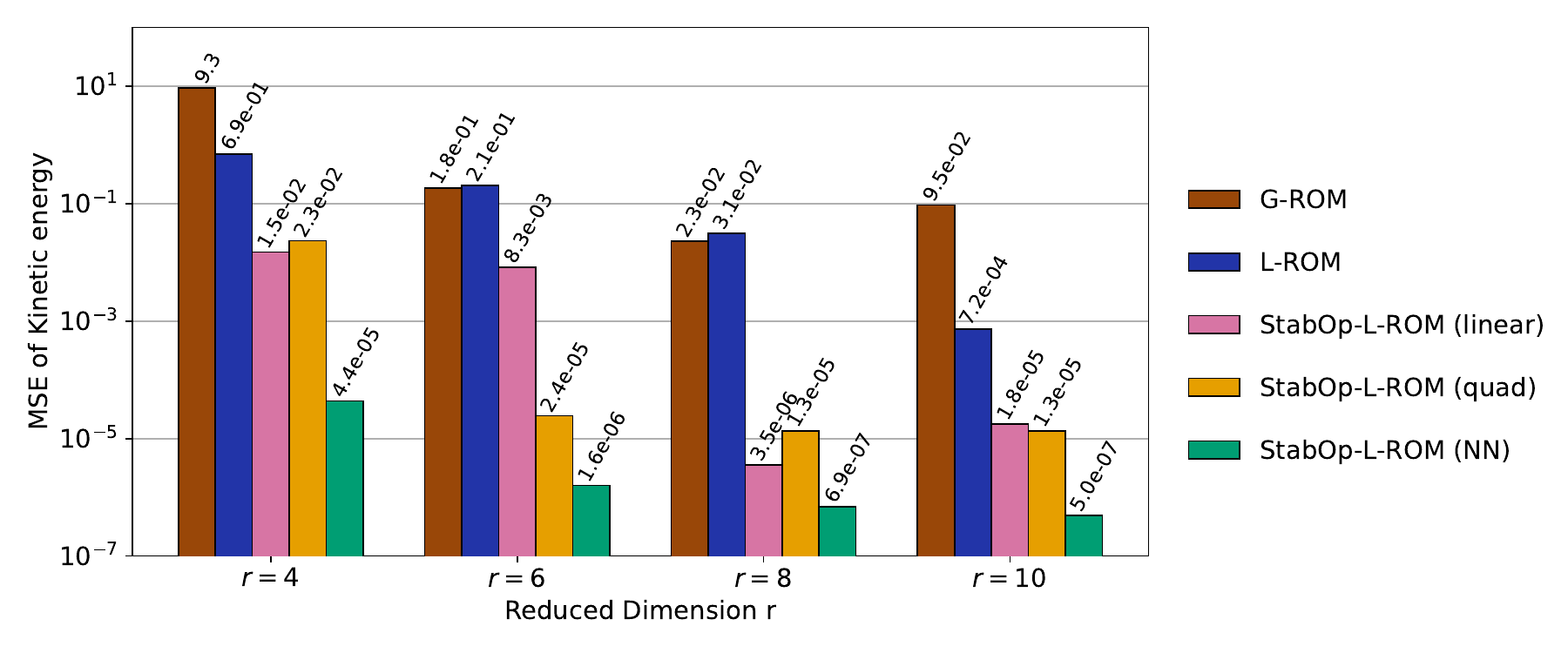}
    \caption{2D flow past a cylinder at $\rm Re=500$. Mean squared error of kinetic energy with respect to the FOM reference for 
    G-ROM, L-ROM, and new {\dlrom}.}
    \label{fig:mse_cyl}
\end{figure}

The accuracy of StabOp-L-ROM is also quantitatively confirmed by Fig.~\ref{fig:mse_cyl}, which shows the mean squared error of the ROM kinetic energy with respect to the FOM over the entire target interval.
All three {\dlrom}s provide 
increased accuracy over the G-ROM and the L-ROM. In particular, {\dlrom} (NN) is the best performing method for all the $r$ values, and it improves the accuracy of {\dlrom} (linear) and {\dlrom} (quadratic) by three orders of magnitude for $r=4$.

\begin{figure}[!ht]
    \centering
    \includegraphics[width=0.8\linewidth]{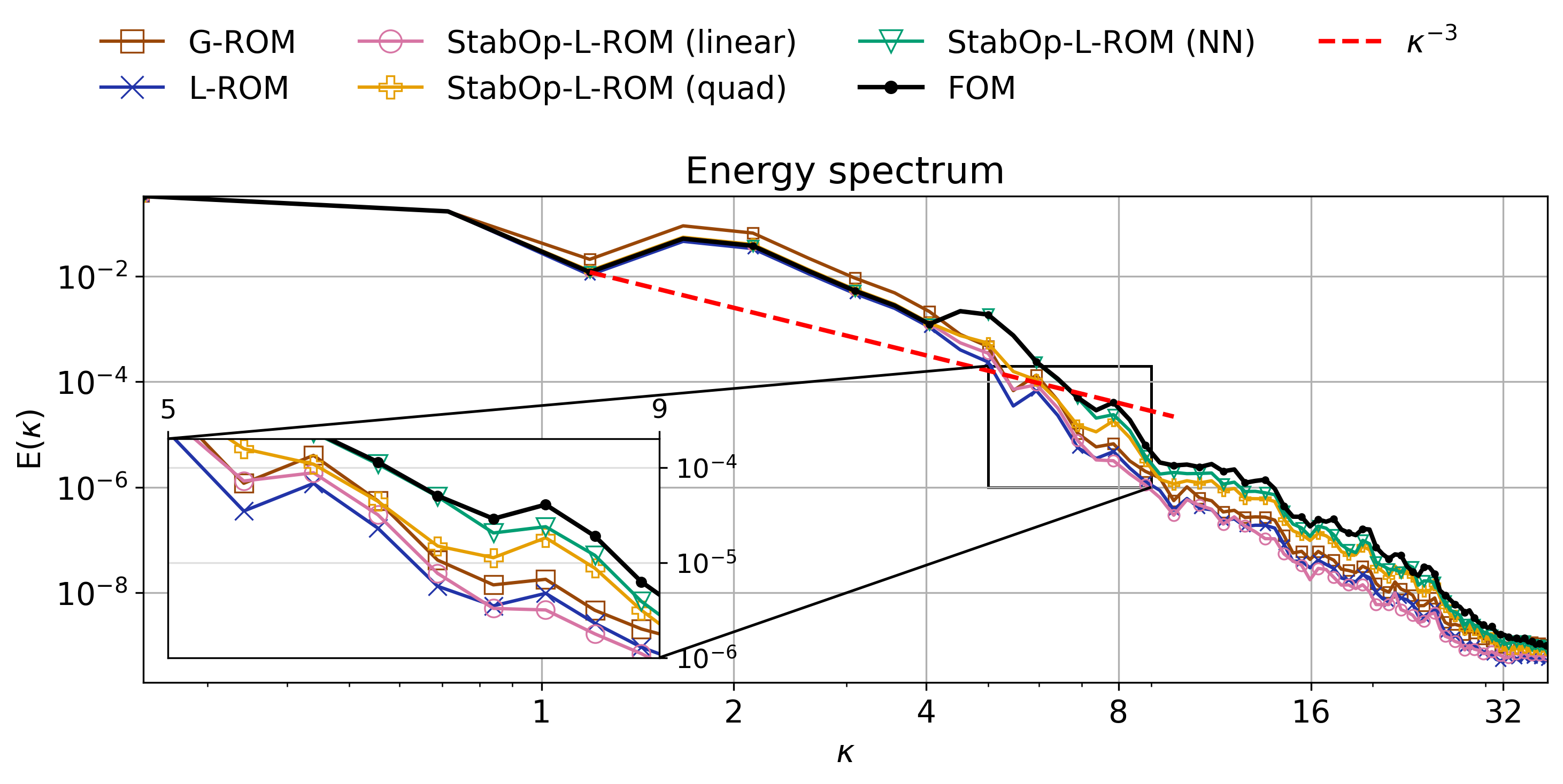}
    \caption{2D flow past a cylinder at $\rm Re=500$. Energy spectrum for FOM, G-ROM, L-ROM, and StabOp-L-ROMs with ROM dimension $r=4$.}
    \label{fig:spectrum-cyl}
\end{figure}

Finally, Fig.~\ref{fig:spectrum-cyl} displays the spatial energy spectrum $E(\kappa)$ at the last time instance of the predictive regime for $r=4$. Since the presence of the cylinder prevents the direct use of FFT-based techniques, the spectrum is computed over the downstream subregion $[1,17]\times[-5,5]\subset\Omega$.
The energy spectrum shows that the G-ROM overestimates the energy at small wavenumbers (large scales), 
which is consistent with the rapid growth of the G-ROM kinetic energy observed in Fig.~\ref{fig:2dfpc_ke}. In contrast, the L-ROM is overly diffusive, as indicated by the underestimated energy at medium and large wavenumbers. The StabOp-L-ROMs show progressively better agreement with
the FOM reference as the StabOp complexity 
increases. In particular, the StabOp-L-ROM (NN) closely matches the FOM, especially at the medium wavenumbers. We note that the spectrum is shown on a logarithmic scale. 
As highlighted in the zoomed-in box in Fig.~\ref{fig:2dfpc_ke}, the StabOp-L-ROM (NN) outperforms the L-ROM by approximately one order of magnitude at the medium scales.

\subsection{2D Lid-Driven Cavity}
\label{results-ldc}

We next consider a more challenging test: the unsteady lid-driven cavity problem at $\rm Re = {10000}$ subject to the steady boundary condition \cite{kaneko2020towards}
\begin{align}
    \begin{cases}
        \bu = [(1-x^2)^2, 0]^T 
        \quad \text{on } \Gamma_{\mathrm{top}}\times \mathbb{R}_+,\\
        \bu = \mathbf{0} \quad \text{on } \partial \Omega \backslash \Gamma_{\mathrm{top}} \times \mathbb{R}_+,\\
        \bu = \mathbf{0} \quad \text{on } \Omega \times \{0\},
    \end{cases}
\end{align}
where 
$\Omega = [-1, 1]^2$ and $\Gamma_{\mathrm{top}} = \{\bx \in \overline{\Omega}: y = 1 \}$. 
The FOM simulation is carried out using Nek5000 \cite{fischer2008nek5000} with
$1024$ spectral elements of polynomial order $N = 7$. We focus on the time interval $[5500,~6300]$, after the solution reaches a statistically steady state region.

\begin{figure}[!ht]
    \centering
    \includegraphics[width=1\textwidth]{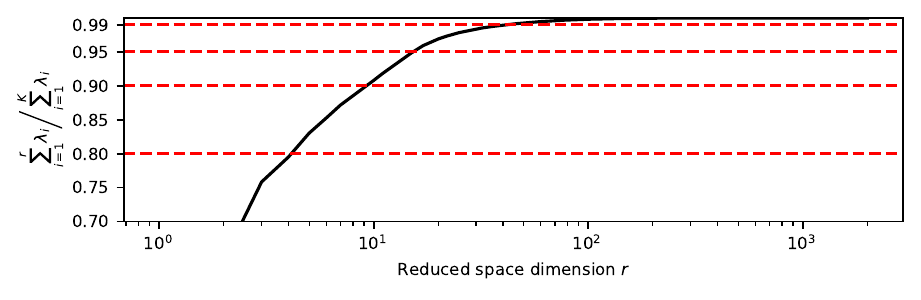}
   \caption{2D lid-driven cavity at $\rm Re={10000}$. The behavior of $\sum^{r}_{i=1} \lambda_i /\sum^{\Ntrain}_{i=1} \lambda_i$ as a function of the reduced space dimension $r$.}
   \label{fig:2dldc_emr}
\end{figure}

The ROM basis functions $\{\bphi_i\}^r_{i=1}$ are constructed via POD using 
$\Ntrain=2001$ snapshots from the time interval $\Ttrain = [5500,~5900]$, corresponding to a sampling frequency of $0.2$. The zeroth mode $\bphi_0$ is defined to be the
mean velocity over $\Ttrain$, and the ROM initial condition is obtained by projecting the lifted snapshot at $t=5500$ onto the ROM space. 
We consider four ROM space dimensions, $r=5, 10, 16, 42$. 
The selected values correspond to energy thresholds $\delta_\sigma = 0.8$, $0.9$, $0.95$, and $0.99$, respectively, as shown in Fig.~\ref{fig:2dldc_emr}. This figure highlights the complexity of the lid-driven cavity flow vs the 2D flow past a cylinder: in the flow past a cylinder test case, with $r=10$ one captures over $99\%$ of the energy, whereas the lid-driven cavity flow exhibits richer dynamics, requiring $r=42$ modes to capture $99\%$ of the energy.

The setup of the {\dlrom}s is reported in Table \ref{tab:data-ldc}.
To improve numerical stability and sensitivity during optimization, the loss values are transformed using a logarithmic function, as they are typically small in magnitude. 
For the linear and quadratic model forms, the matrix $\tA$ is initialized using the differential filter associated with a given $\delta$ (see Remark~\ref{remark:initialization}), while the bias term $\uwb$ is omitted. Preliminary tests indicated that including $\uwb$ degraded the model performance in both cases.
For the nonlinear model form, the differential filter is incorporated directly into the network architecture (see Remark~\ref{remark:initialization}). As indicated in the last column of Table~\ref{tab:data-ldc}, the filter radius is treated as an additional hyperparameter in all StabOp training procedures, with the search space $\mP_{\delta}$ consisting of $38$ logarithmically spaced values between $0.001$ and $0.4$. 
\begin{table}[htpb!]
    \centering
    \begin{tabular}{>{\centering\arraybackslash}m{0.4cm} 
                    >{\centering\arraybackslash}m{1.2cm} 
                    >{\centering\arraybackslash}m{1.2cm} 
                    >{\centering\arraybackslash}m{1.cm}
                    >{\centering\arraybackslash}m{1.cm}
                    >{\centering\arraybackslash}m{1.cm}
                    >{\centering\arraybackslash}m{0.6cm}
                    >{\centering\arraybackslash}m{1cm}
                    >{\centering\arraybackslash}m{0.8cm}
                    >{\centering\arraybackslash}m{1.9cm}
                    }
    \toprule
\multirow{2}{*}{$r$}&\multirow{2}{1.2cm}{Energy retained}&\multicolumn{3}{c}{Time window}&\multirow{2}{*}{$h$ (\texttt{NN})}&\multirow{2}{*}{$\eta$}&\multirow{2}{*}{$\Nepoch$}&\multirow{2}{*}{$\alpha$} 
&\multirow{2}{*}{$\delta$}\\
\cmidrule{3-5}
&& $\Ttrain$ & $\Tval$ & $\Ttest$ &&&&&\\
\midrule
5&$0.8$&\multirow{5}{1cm}{$[5500, \allowbreak 5900]$}&\multirow{5}{1cm}{$[5900,\allowbreak 6100]$}&\multirow{5}{1cm}{$[6100, \allowbreak 6300]$}&Tanh&\multirow{5}{*}{$0.2$}&\multirow{5}{*}{$200$}&\multirow{5}{*}{$10^{-8}$}
&\multirow{5}{1.7cm}{optimized in $\mathcal{P_{\delta}} \subset [0.001, 0.4]$}\\
\cmidrule{1-2}\cmidrule{6-6}
10&$0.9$&&&&Tanh&&&&\\
\cmidrule{1-2}\cmidrule{6-6}
16&$0.95$&&&&ReLU&&&&\\
\cmidrule{1-2}\cmidrule{6-6}
42&$0.99$&&&&ReLU&&&&\\
    \bottomrule
    \end{tabular}
    \caption{2D lid-driven cavity at $\rm Re={10000}$. Hyperparameters for the {\dlrom}s (linear, quadratic, and fully nonlinear). 
    Columns labeled \texttt{NN} correspond to hyperparameters specific to the neural-network model, namely the activation function.}
    \label{tab:data-ldc}
\end{table}
For L-ROM, the filter radius hyperparameter space described at the beginning of this section is considered.

\begin{figure}[!ht]
    \centering
    \includegraphics[width=0.9\textwidth]{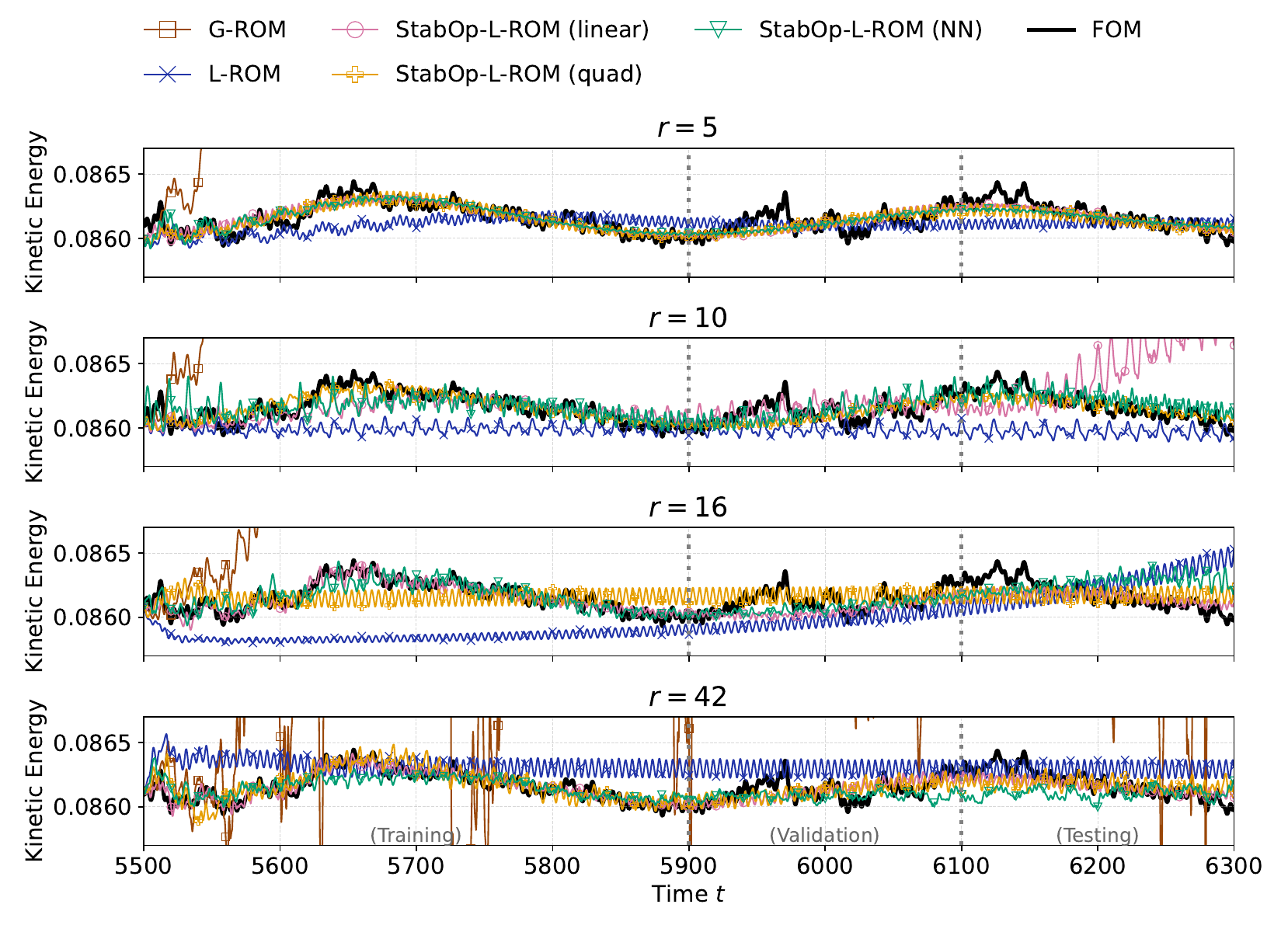}
    
   \caption{2D lid-driven cavity at $\rm Re={10000}$. Kinetic energy behavior of the {\dlrom}, along with the results of the G-ROM, L-ROM with an optimal filter radius, and FOM.}
   \label{fig:2dldc_ke}
\end{figure}

Fig.~\ref{fig:2dldc_ke} shows the kinetic energy evolution on the target interval $[5500,~6300]$ for the four ROM dimensions investigated. The training, validation, and test intervals are indicated in the figure by vertical dotted lines. The comparison includes five ROMs, that is, G-ROM, L-ROM, and {\dlrom} with linear, quadratic, and nonlinear model forms, along with the FOM. 

For all values of $r$, G-ROM is unstable and fails to provide accurate long-time predictions, displaying significant deviations from the FOM. 
With the optimal filter radius, L-ROM yields stable predictions for $r=5, 10$, and $42$, but the kinetic energy displays inaccurate periodic or quasi-periodic behavior. For $r=16$, the energy is initially stabilized but gradually grows in time. 

In contrast, the proposed {\dlrom}s show significantly better results compared to the L-ROM. Indeed, the {\dlrom} (linear) captures the FOM kinetic behavior over the target interval, with the exception of $r=10$, where the energy increases in the test interval. 
The {\dlrom} (quad) performs well across all cases except for $r = 16$, where it produces a stable prediction that slightly underestimates the FOM energy but captures the correct mean kinetic energy. The {\dlrom} (NN) performs well across all cases except for $r=42$, where it is less accurate than {\dlrom} (linear) and {\dlrom} (quad).

\begin{figure}[htpb!]
    \centering
    \includegraphics[width=\linewidth]{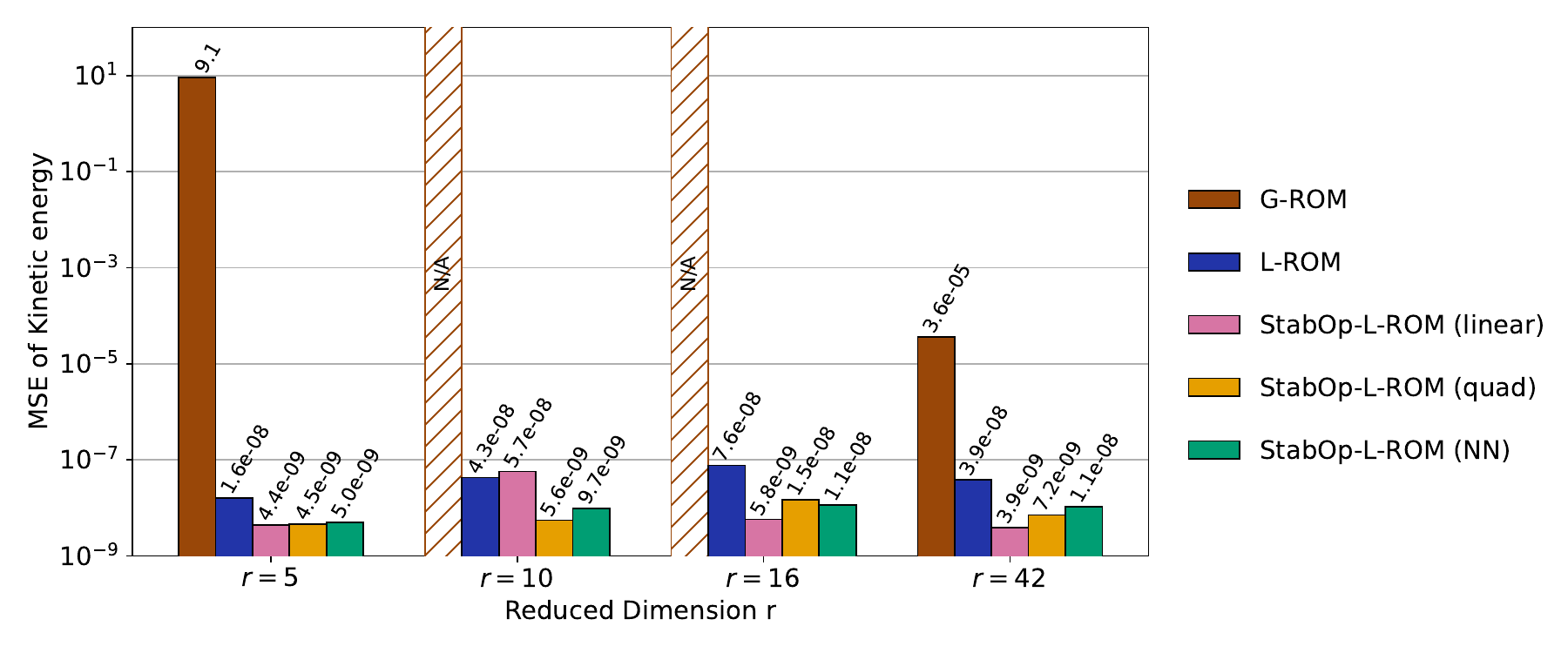}
    \caption{2D lid-driven cavity at $\rm Re={10000}$. Mean squared error of kinetic energy with respect to the FOM reference for G-ROM, L-ROM, and {\dlrom} with linear, quadratic, and nonlinear model forms.}
    \label{fig:mse_ldc}
\end{figure}

To further quantify the differences among the {\dlrom}s, Fig.~\ref{fig:mse_ldc} reports the mean squared error of the ROM kinetic energy with respect to the FOM, for different $r$ values and for all models considered. Fig.~\ref{fig:mse_ldc} confirms that the G-ROM is highly unstable. 
With the exception of $r=10$, all {\dlrom}s consistently outperform the classical L-ROM equipped with an optimal filter in terms of the MSE of the kinetic energy. 
For $r=10$, the {\dlrom} (linear) is slightly less accurate than the L-ROM, due to the energy growth in the testing interval. Overall, these results demonstrate that the proposed {\dlrom} substantially improves both stability and accuracy compared to the L-ROM.

\begin{figure}[htpb!]
    \centering
    \includegraphics[width=0.8\linewidth]{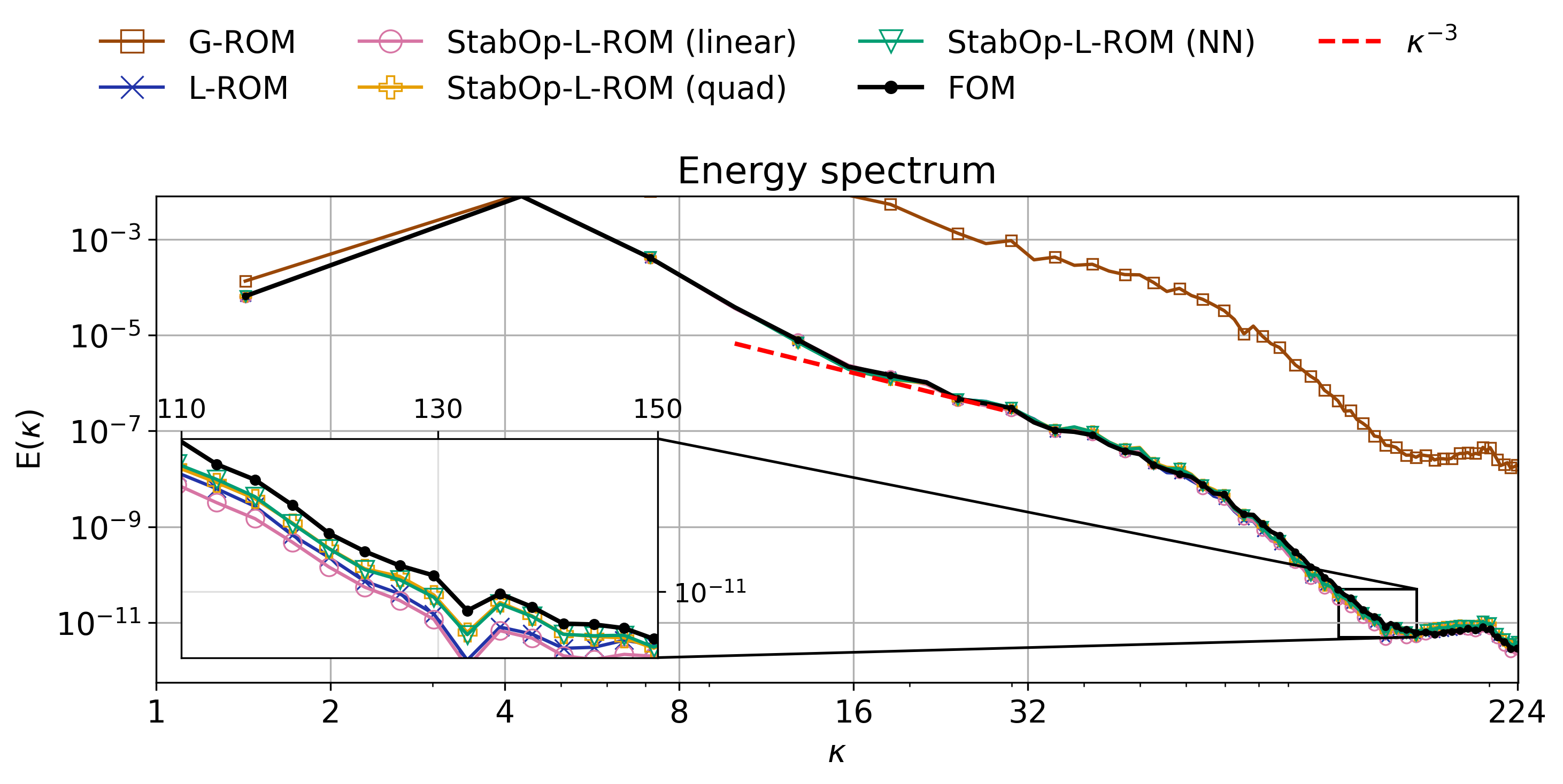}
    \caption{2D lid-driven cavity at $\rm Re={10000}$. Energy spectrum for FOM, G-ROM, L-ROM, and StabOp-L-ROMs 
    with ROM dimension $r=5$.}
    \label{fig:spectrum-ldc}
\end{figure}

Finally, Fig.~\ref{fig:spectrum-ldc} displays the energy spectrum $E(\kappa)$ at the last time instance of the predictive regime for $r=5$. The energy is computed considering the entire spatial domain.
The spectrum shows that the G-ROM overestimates the energy at all wavenumbers, 
which is consistent with the uncontrolled growth of the G-ROM kinetic energy observed in Fig.~\ref{fig:2dldc_ke}.
In contrast, the L-ROM and StabOp-L-ROMs 
closely match the reference FOM. 
At this final time, the StabOp energy spectrum does not exhibit a clear improvement. This observation is consistent with Fig.~\ref{fig:2dldc_ke}, where the L-ROM and StabOp-L-ROMs yield similar predictions at the final time while differing at intermediate time instances.

\subsection{3D Flow Past a Hemisphere}
\label{results-hemi}

We next consider the 3D flow past a hemisphere at $\rm Re={2200}$ \cite{tufo1999numerical}. The computational domain is a rectangular channel with depth $3.25$ (along the $z$-axis), width $6.4$ (along the $y$-axis), and length
$18.2$ (along the $x$-axis). The hemisphere, with radius $0.5$, is mounted on a thin cylindrical base of radius $0.5$ and height $0.05$, which is placed on a smooth flat plate at $z= 0$.

Inflow and outflow boundaries are imposed in the $x$-direction. 
At the inlet, a Dirichlet boundary condition is prescribed using a smooth, sine-based inflow profile that approximates a boundary layer structure:
\begin{align}
u(z) =
\begin{cases}
\sin\left( \dfrac{\pi z}{1.2} \right), & z \le 0.6, \\[6pt]
1, & z > 0.6.
\end{cases}
\end{align}
Here, $0.6$ is the boundary layer thickness. 
This inflow condition is not the exact Blasius solution but serves as a smooth analytic approximation that captures the key transition from zero velocity at the wall to the free-stream value away from the wall. 
At the outlet, a homogeneous Neumann boundary condition is imposed for the velocity. Symmetry boundary conditions are applied on the lateral boundaries and at the top wall  ($z = 3.25$), enforcing zero normal velocity and zero tangential shear stress. A no-slip boundary condition is imposed at the bottom wall 
($z=0$). 

The FOM simulation is carried out using nekRS \cite{fischer2022nekrs} with
$2042$ spectral elements of polynomial order $N = 9$, resulting in about $2 \times 10^6$ degrees of freedom.
We focus on the time interval $[3000,~3200]$, after the solution reaches a statistically steady state region.

\begin{figure}[!ht]
    \centering
    \includegraphics[width=1\textwidth]{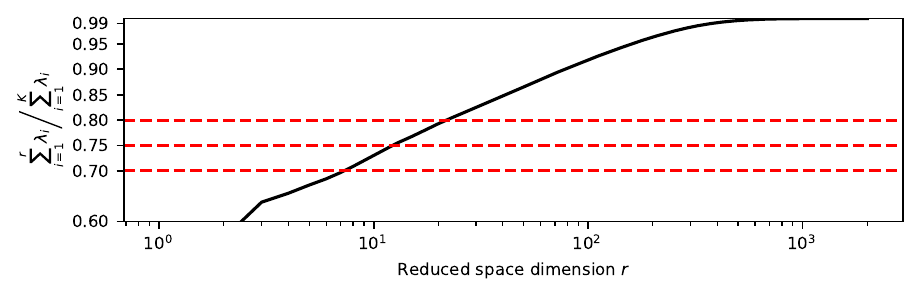}
    
   \caption{3D flow past a hemisphere at $\rm Re={2200}$. The behavior of $\sum^{r}_{i=1} \lambda_i /\sum^{\Ntrain}_{i=1} \lambda_i$ as a function of the reduced space dimension, $r$.}
   \label{fig:hemi_efr}
\end{figure}

The reduced basis functions $\{\bphi_i\}^r_{i=1}$ are constructed via 
POD from $\Ntrain={2001}$ snapshots collected over the time interval $\Ttrain  = [{3000},~ {3100}]$, corresponding to a sampling frequency of $0.05$. The zeroth mode $\bphi_0$ is defined as the velocity at $t={3000}$, and the ROM initial condition is 
obtained by projecting the lifted snapshot at $t={3000}$ onto the reduced space. 
We consider three reduced space dimensions, $r=7, 12, 20$, which are determined based on the energy criterion (\ref{eq:energy_criteria}) and correspond to energy thresholds $\delta_\sigma = 0.7, 0.75, 0.8$, respectively, as shown in Fig.~\ref{fig:hemi_efr}. 

The StabOp, $\mF$, in {\dlrom} is trained using the hyperparameters listed in Table \ref{tab:data-hemi}. 
\begin{table}[htpb!]
    \centering
    \begin{tabular}{>{\centering\arraybackslash}m{0.5cm} 
                    >{\centering\arraybackslash}m{1.2cm} 
                    >{\centering\arraybackslash}m{1.cm} 
                    >{\centering\arraybackslash}m{1.cm}
                    >{\centering\arraybackslash}m{1.cm}
                    >{\centering\arraybackslash}m{0.8cm}
                    >{\centering\arraybackslash}m{0.6cm}
                    >{\centering\arraybackslash}m{1cm}
                    >{\centering\arraybackslash}m{0.9cm}
                    >{\centering\arraybackslash}m{1.9cm}
                    >{\centering\arraybackslash}m{1.9cm}
                    }
    \toprule
\multirow{2}{*}{$r$}&\multirow{2}{1.2cm}{Energy retained}&\multicolumn{3}{c}{Time window}&\multirow{2}{*}{$h$ (\texttt{NN})}&\multirow{2}{*}{$\eta$}&\multirow{2}{*}{$\Nepoch$}&\multirow{2}{*}{$\alpha$} 
&\multirow{2}{*}{$\delta$}\\
\cmidrule{3-5}
&& $\Ttrain$ & $\Tval$ & $\Ttest$ &&&&&\\
\midrule
7&$0.7$&\multirow{4}{1cm}{$[3000, \allowbreak 3100]$}&\multirow{4}{1cm}{$[3100,\allowbreak 3150]$}&\multirow{4}{1cm}{$[3150, \allowbreak 3200]$}&\multirow{4}{*}{SiLU}&\multirow{4}{*}{$0.2$}&\multirow{4}{*}{$200$}&\multirow{4}{*}{$10^{-8}$}
&\multirow{4}{1.7cm}{optimized in $\mathcal{P_{\delta}} \subset [0.001, 0.1]$}\\
\cmidrule{1-2}
12&$0.75$&&&&&&&\\
\cmidrule{1-2}
20&$0.8$&&&&&&&\\
    \bottomrule
    \end{tabular}
    \caption{3D flow past a hemisphere at $\rm Re={2200}$. Hyperparameters for the {\dlrom}s (linear, quadratic, and fully nonlinear). 
    Columns labeled \texttt{NN} correspond to hyperparameters specific to the neural-network model, namely the activation function.}
    \label{tab:data-hemi}
\end{table}
For the linear and quadratic forms, the matrix $\tA$ is initialized using the differential filter associated with a given $\delta$, as described in Remark~\ref{remark:initialization}. In contrast, for the nonlinear model form, the differential filter is incorporated directly into the model formulation, as discussed in the same remark. The filter radius of the differential filter is treated as an additional hyperparameter, as shown in Table \ref{tab:data-hemi}, and is optimized among $10$ logarithmically spaced values between $0.001$ and $0.1$.
For L-ROM, the filter radius hyperparameter space described at the beginning of this section is considered.

\begin{figure}[!ht]
    \centering
    \includegraphics[width=0.9\textwidth]{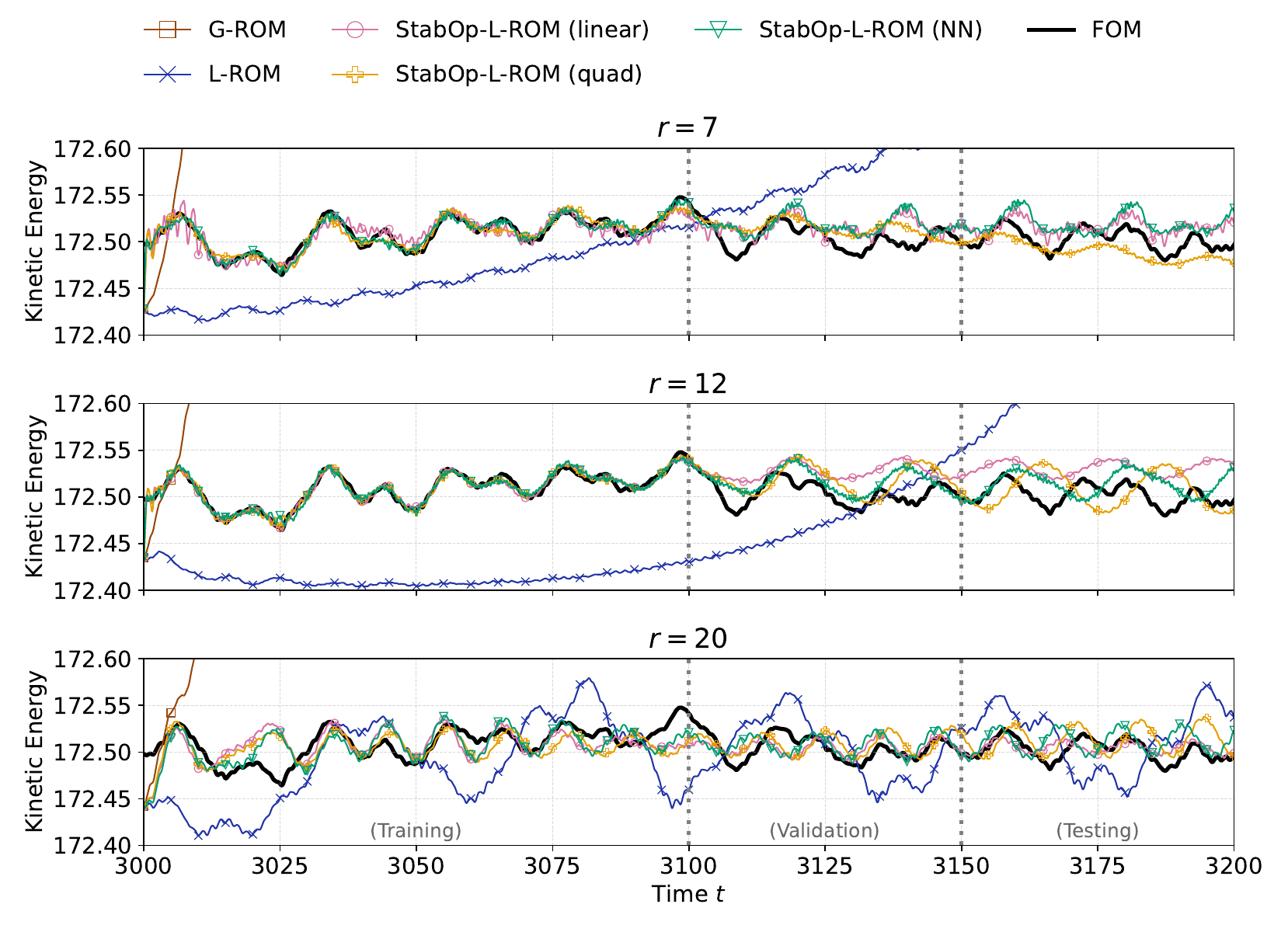}
    \caption{3D flow past a hemisphere at $\rm Re={2200}$. Kinetic energy behavior of the {\dlrom}, along with the results of the G-ROM, L-ROM with an optimal filter radius, and FOM.}
    \label{fig:hemi_ke}
\end{figure}

Fig.~\ref{fig:hemi_ke} shows the kinetic energy evolution on the target interval $[3000, ~3200]$ for the three ROM dimensions investigated. The training, validation, and test intervals are indicated in the figure by vertical dotted lines. The comparison includes five ROMs, that is, G-ROM, L-ROM, and {\dlrom} with linear, quadratic, and nonlinear model forms, along with the FOM. 

\begin{figure}[htpb!]
    \centering
    \includegraphics[width=\linewidth]{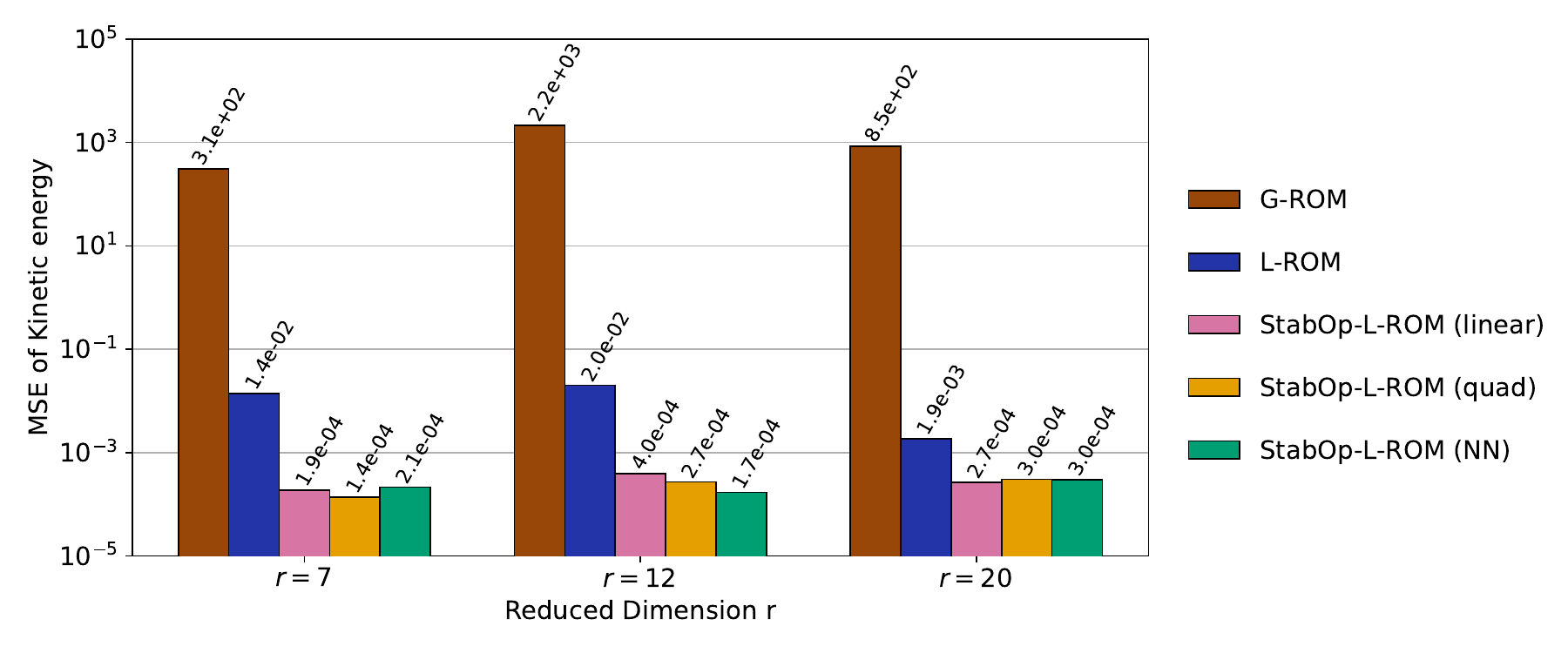}
    \caption{3D flow past a hemisphere at $\rm Re={2200}$. Mean squared error of kinetic energy with respect to the FOM reference for G-ROM, L-ROM, and {\dlrom} with linear, quadratic, and nonlinear model forms.}
    \label{fig:mse_hemi}
\end{figure}

For all values of $r$, G-ROM is unstable and fails to provide accurate long-time predictions, displaying significant deviations from the FOM right at the beginning of the simulation. 
With the optimal filter radius, the predicted kinetic energy of L-ROM grows in time for $r=7$ and $r=12$. For $r=20$, the L-ROM predicted kinetic energy is stable but has a much larger oscillation compared to the FOM.
In contrast, all the {\dlrom}s show significantly better results compared to the L-ROM equipped with an optimal filter radius. 
For $r = 7$ and $r = 12$, all three 
{\dlrom}s accurately reconstruct the FOM kinetic energy over $\Ttrain$ and exhibit periodic-like behavior in the validation and testing intervals. Although none of the variants fully reconstruct 
the kinetic energy in $\Ttrain$ at $r = 20$, all three 
{\dlrom}s yield accurate predictions and remain stable in the validation and testing intervals. Overall, these results demonstrate that the proposed {\dlrom} substantially improves both stability and accuracy of the classical L-ROM equipped with an optimal filter radius. This is also confirmed in Fig.~\ref{fig:mse_hemi}, which quantifies the mean squared error of the kinetic energy with respect to the FOM, for all models considered and for all the ROM dimensions.
While the optimized L-ROM is more accurate than the G-ROM,
its error is still large. All three {\dlrom}s yield similar accuracy and improve the accuracy of the L-ROM by one or two orders of magnitude.

\begin{figure}[htpb!]
    \centering
    \includegraphics[width=0.8\linewidth]{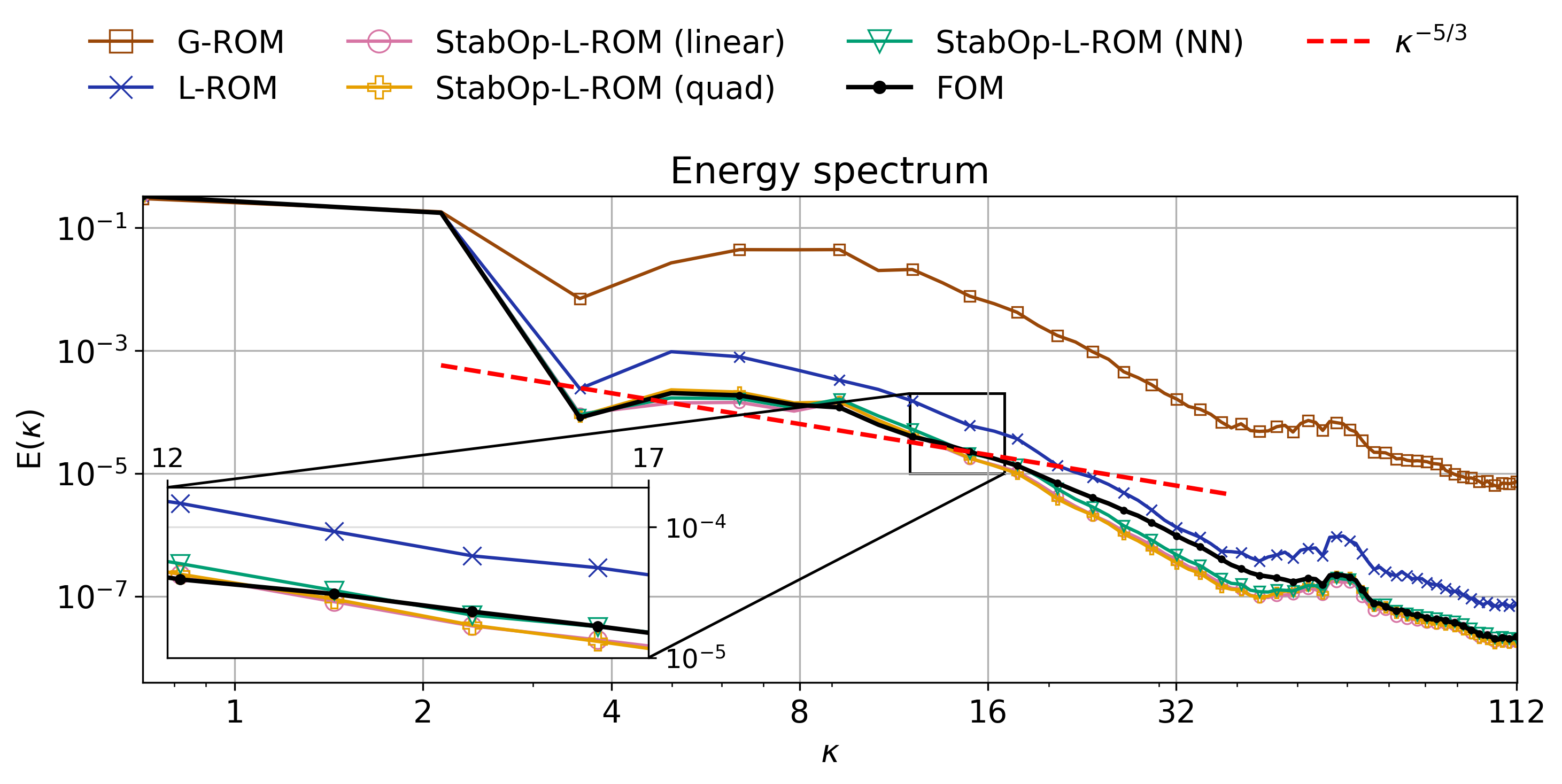}
    \caption{3D flow past a hemisphere at $\rm Re={2200}$. Energy spectrum for FOM, G-ROM, L-ROM, and StabOp-L-ROMs 
    with ROM dimension $r=12$.}
    \label{fig:spectrum-hemi}
\end{figure}

Finally, Fig.~\ref{fig:spectrum-hemi} shows the energy spectrum $E(\kappa)$ at the last time instance of the predictive regime for $r=12$. 
As in the flow past a cylinder test case, the spectrum is computed over a region downstream of the hemisphere, given by $[2, 15] \times [-3.2, 3.2] \times [0, 3.25]$.
The spectrum indicates that both the G-ROM and the L-ROM overestimate the FOM energy across almost all wavenumbers, which is consistent with the uncontrolled energy growth observed in Fig.~\ref{fig:hemi_ke}. 
In contrast, all StabOp-L-ROMs closely match the reference FOM at the last time instance, in agreement with the results shown in Fig.~\ref{fig:hemi_ke}.

\subsection{Minimal Channel Flow}
\label{results-minimal}

We consider the minimal channel flow at $\rm Re={5000}$, which presents strong turbulent features 
while maintaining simplified flow dynamics, resulting in significantly lower computational costs compared to a full channel flow simulation
\cite{jimenez1991minimal}. Following the setup in
\cite{jimenez1991minimal}, the streamwise and spanwise lengths of the channel are set to $0.6 \pi$ and $0.18\pi$, respectively, and the channel
half-height is set to $1$. 

The FOM simulation is carried out using nekRS \cite{fischer2022nekrs} with $576$ spectral elements of polynomial order $N = 9$, resulting in about $5.76 \times 10^5$ degrees of freedom. We focus on the time interval $[3000,~4000]$, after the solution reaches a statistically steady state region. {This represents a challenging test problem for reduced order modeling, since} G-ROM requires $r \ge 400$ to capture the kinetic energy of the FOM \cite{tsai2023accelerating}.

\begin{figure}[!ht]
    \centering
    \includegraphics[width=0.9\textwidth]{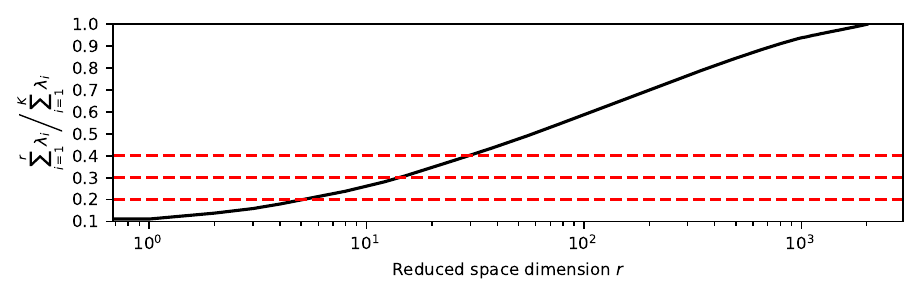}
    
    \caption{3D minimal channel flow at $\rm Re={5000}$. The behavior of $\sum^{r}_{i=1} \lambda_i /\sum^{\Ntrain}_{i=1} \lambda_i$ as a function of the ROM space dimension, $r$.}
    \label{fig:mfu_efr}
\end{figure}

The ROM basis functions $\{\bphi_i\}^r_{i=1}$ are constructed via POD from $\Ntrain={2001}$ snapshots collected over the time interval $\Ttrain  = [{3000},~ {3500}]$, corresponding to a samplig frequence of $0.25$. The zeroth mode,  $\bphi_0$, is set to the mean velocity over $\Ttrain$, and the ROM initial condition is obtained by projecting the lifted snapshot at $t={3000}$ onto the ROM space. 
We consider three ROM space dimensions, $r=6, 15, 30$, which are determined based on the energy criterion (\ref{eq:energy_criteria}) and correspond to the energy thresholds $\delta_\sigma = 0.2, 0.3, 0.4$, respectively, as shown in Fig.~\ref{fig:mfu_efr}.

The StabOp, $\mF$, in {\dlrom} is trained using the hyperparameters listed in Table \ref{tab:data-mfu}.
To improve numerical stability and sensitivity during optimization, the loss values are transformed using a logarithmic function, as they are typically small in magnitude. 
\begin{table}[htpb!]
    \centering
    \begin{tabular}{>{\centering\arraybackslash}m{0.5cm} 
                    >{\centering\arraybackslash}m{1.2cm} 
                    >{\centering\arraybackslash}m{1.cm} 
                    >{\centering\arraybackslash}m{1.cm}
                    >{\centering\arraybackslash}m{1.cm}
                    >{\centering\arraybackslash}m{0.8cm}
                    >{\centering\arraybackslash}m{0.6cm}
                    >{\centering\arraybackslash}m{1cm}
                    >{\centering\arraybackslash}m{0.9cm}
                    >{\centering\arraybackslash}m{2.5cm}
                    }
    \toprule
\multirow{2}{*}{$r$}&\multirow{2}{1.2cm}{Energy retained}&\multicolumn{3}{c}{Time window}&\multirow{2}{*}{$h$ (\texttt{NN})}&\multirow{2}{*}{$\eta$}&\multirow{2}{*}{$\Nepoch$}&\multirow{2}{*}{$\alpha$}&\multirow{2}{*}{$\delta$ (\texttt{linear}, \texttt{quad})}\\
\cmidrule{3-5}
&& $\Ttrain$ & $\Tval$ & $\Ttest$ &&&&&\\
\midrule
6&$0.2$&\multirow{4}{1cm}{$[3000, \allowbreak 3500]$}&\multirow{4}{1cm}{$[3500,\allowbreak 3750]$}&\multirow{4}{1cm}{$[3750, \allowbreak 4000]$}&\multirow{4}{*}{SiLU}&\multirow{4}{*}{$0.2$}&\multirow{4}{*}{$200$}&\multirow{4}{*}{$10^{-8}$}&{optimized in $\mathcal{P_{\delta}} \subset [0.08, 0.3]$}\\
\cmidrule{1-2}\cmidrule{10-10}
15&$0.3$&&&&&&&&\multirow{2}{2.4cm}{optimized in $\mathcal{P_{\delta}} \subset [0.03, 0.2]$}\\
\cmidrule{1-2}
30&$0.4$&&&&&&&\\
    \bottomrule
    \end{tabular}
    \caption{3D minimal channel flow at $\rm Re={5000}$. Hyperparameters for the {\dlrom}s (linear, quadratic, and fully nonlinear).
    Columns labeled \texttt{linear}, \texttt{quad}, and \texttt{NN} correspond to hyperparameters specific to the linear, quadratic or neural-network model, respectively.}
    \label{tab:data-mfu}
\end{table}

For the linear and quadratic forms, instead of initializing the matrix $\tA$ as the differential filter associated with a given $\delta$, as done in the previous test cases, we incorporate the differential filter into the model formulation following the strategy described in Remark~\ref {remark:initialization}. We omit the bias term $\uwb$ in both linear and quadratic forms, since preliminary tests indicated that including $\uwb$ degraded the model performance in both cases. 
For the nonlinear model form, the differential filter is incorporated directly into the model formulation, as discussed in the same remark.

\begin{figure}[!ht]
    \centering
    \includegraphics[width=0.85\textwidth]{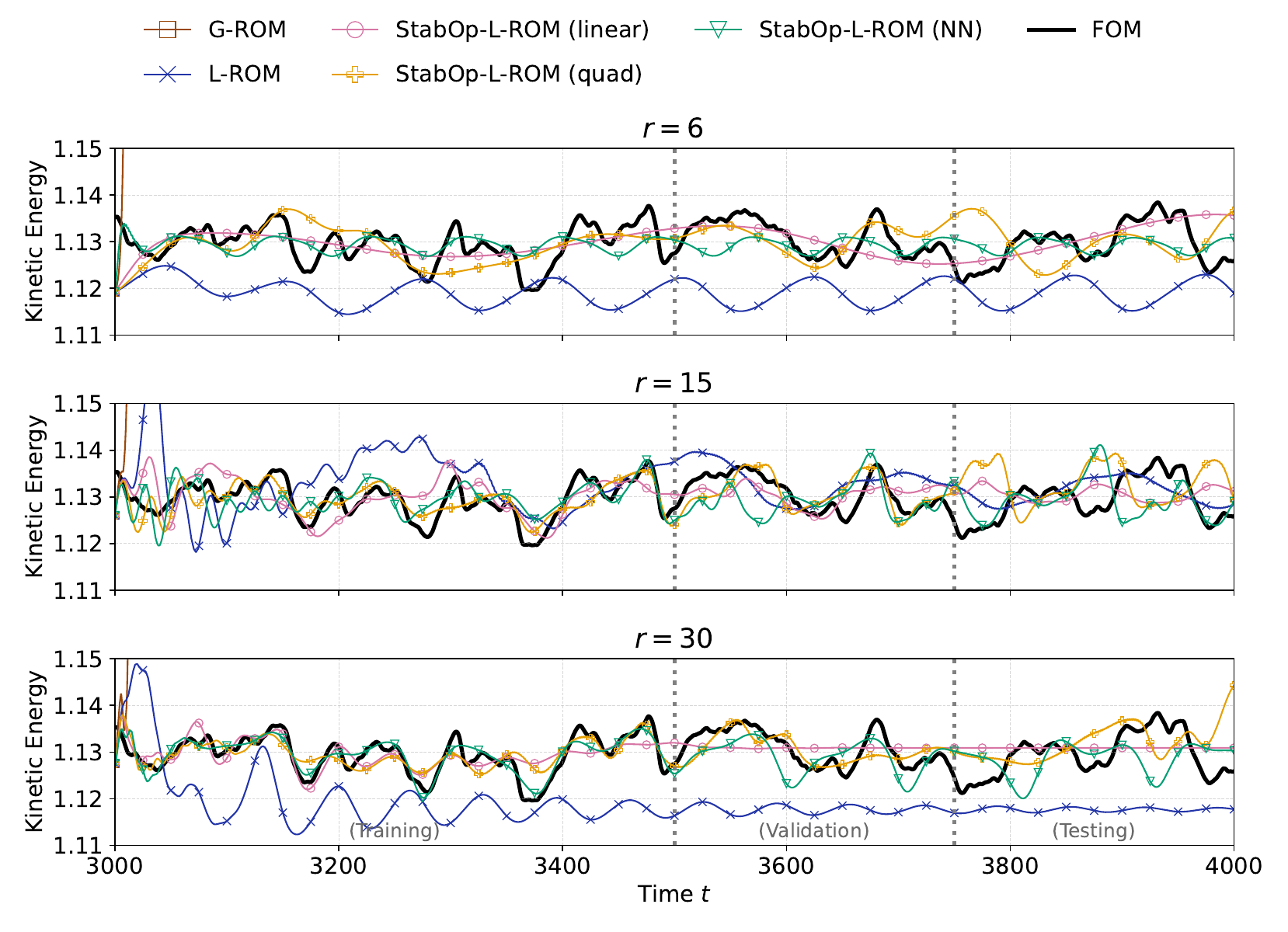}
   \caption{3D minimal channel flow at $\rm Re={5000}$. Kinetic energy behavior of the {\dlrom}, along with the results of the G-ROM, L-ROM with an optimal filter radius, and FOM.}
   \label{fig:mfu_ke}
\end{figure}

Figure~\ref{fig:mfu_ke} shows the kinetic energy evolution on the target interval $[3000,~4000]$ for the three ROM dimensions investigated given by the G-ROM, L-ROM, and {\dlrom} with linear, quadratic, and nonlinear model forms. The training, validation, and test intervals are indicated in the figure by vertical dotted lines. 

For all $r$ values, G-ROM is unstable and deviates from the FOM right from the beginning of the simulation.
With the optimal filter radius, L-ROM yields stable predictions for $r=6$, but the kinetic energy exhibits inaccurate periodic behavior. 
For $r=15$, L-ROM produces a stable trajectory that roughly follows the mean trend of the FOM kinetic energy but exhibits large oscillations during the early part of the training interval and fails to capture finer temporal variations. For $r=30$, the energy prediction is more stable but saturates to a steady state, failing to reflect the time-dependent behavior of the FOM.

In contrast, the proposed {\dlrom}, particularly the quadratic and nonlinear variants, shows improved performance in capturing the FOM kinetic energy. For all $r$ values, the {\dlrom} (linear) captures the mean trend of the FOM energy but appears overly smoothed, whereas both the {\dlrom} (quad) and {\dlrom} (NN) better capture the temporal variation of the kinetic energy. 
For $r=15$ and $r=30$, 
all three {\dlrom}s show better alignment with the FOM in terms of both amplitude and temporal variation compared to the $r=6$ case.
For $r=30$, all three {\dlrom}s reproduce the FOM kinetic energy during $\Ttrain$. 
In the validation and testing intervals, the {\dlrom} (linear) becomes overly dissipative, whereas the {\dlrom} (quad) {\dlrom} (NN) variants remain both stable and accurate. 
Overall, these results demonstrate that the proposed {\dlrom} improves both stability and accuracy compared to the L-ROM. 

\begin{figure}[!ht]
    \centering
    \includegraphics[width=0.9\linewidth]{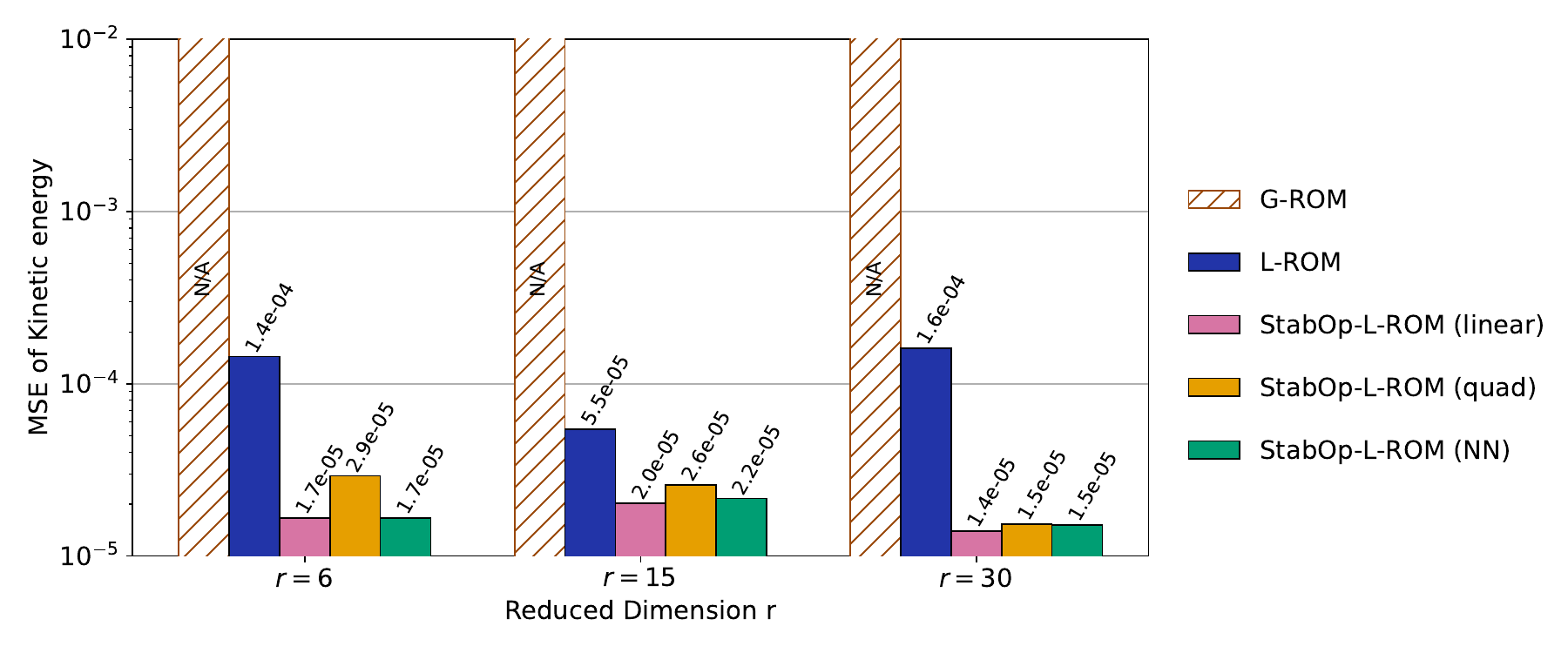}
    \caption{3D minimal channel flow at $\rm Re={5000}$. Mean squared error of kinetic energy with respect to the FOM reference for G-ROM, L-ROM, and StabOp-L-ROM with linear, quadratic, and nonlinear model forms.}
    \label{fig:mse_mfu}
\end{figure}

The improved performance of the {\dlrom} is confirmed by the error analysis in Fig.~\ref{fig:mse_mfu}. The G-ROM exhibits unbounded growth in kinetic energy, 
and thus its mean squared error cannot be quantified (the diagonal slash patterns in Fig.~\ref{fig:mse_mfu}). While the L-ROM yields improved accuracy, it is consistently less accurate than the {\dlrom}s for all considered values of $r$. All three {\dlrom}s yield comparable accuracy in terms of the mean squared error of the kinetic energy.

\begin{figure}[htpb!]
    \centering
    \includegraphics[width=0.8\linewidth]{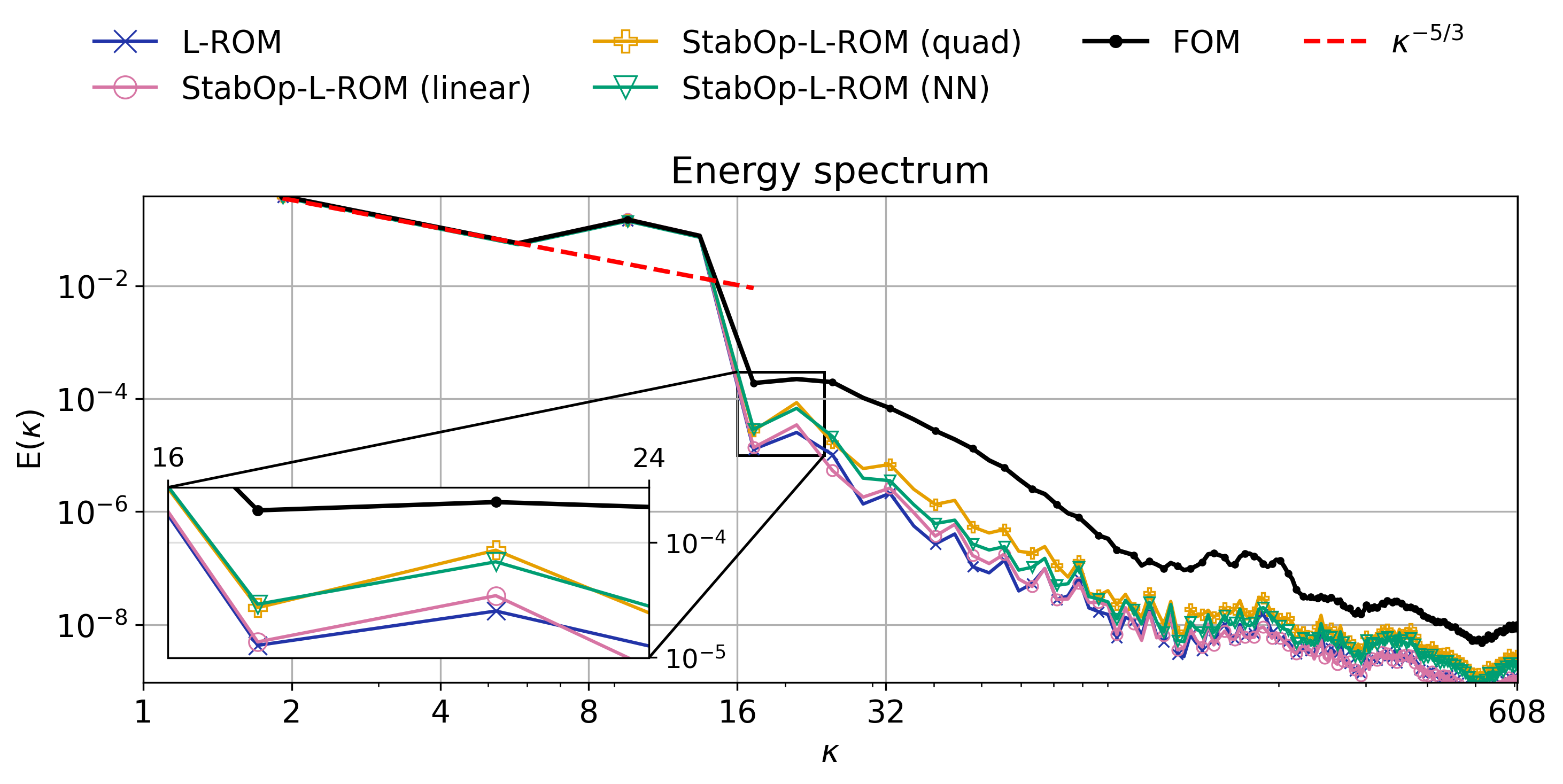}
    \caption{3D minimal channel flow at $\rm Re={5000}$. Energy spectrum for FOM, G-ROM, L-ROM, and StabOp-L-ROMs with ROM dimension $r=6$.
    }
    \label{fig:spectrum-mfu}
\end{figure}

Finally, Fig.~\ref{fig:spectrum-mfu} displays the energy spectrum $E(\kappa)$ at the last time instance of the predictive regime, computed over the entire domain for $r=6$. 
The G-ROM spectrum is not included in the plot, as its energy blows up after the first few time steps. Unlike the other test cases, both the L-ROM and the StabOp-L-ROMs are overly diffusive, as indicated by the underestimated energy at medium and large wavenumbers. 
(We note, however, that this test case is significantly more 
under-resolved than the previous three test cases.) Nevertheless, the StabOp-L-ROMs, particularly the quadratic model, are more accurate than the L-ROM, as further illustrated by the zoomed-in box in Fig.~\ref{fig:spectrum-mfu}.

\subsection{A Comparison of Numerical Stability, Accuracy, and Computational Cost}

In this section, we perform an overall comparison and further assessment of the results obtained for each test case in Sec.~\ref{results-cyl}-\ref{results-minimal} for different models and ROM dimensions. 
In particular, we aim at giving a concise, graphical overview of the overall performance of the new StabOp-L-ROM (in its three variants). 
To this end, we compare StabOp-L-ROM with the standard L-ROM and G-ROM.
Specifically, we first assess the models' stability and accuracy.
Then, we assess the models' offline and online computational cost.

For each test case in Sec.~\ref{results-cyl}-\ref{results-minimal}, we report
(i) the percentage of time each model is \emph{stable}, and 
(ii) the percentage of time each model is \emph{most accurate}.
The percentage of time the model is \emph{stable} is computed as the fraction of time steps for which the model’s kinetic energy remains within a prescribed tolerance relative to the FOM kinetic energy. 
Specifically, at each time step $t$, the ROM energy $\EROM(t)$ is checked against the following limits:
\begin{equation}
    E_{\text{min}}(t)  < \EROM(t) < E_{\text{max}}(t),
\end{equation}
where the bounds are defined as 
\begin{equation}
     E_{\text{min}}(t) = \EFOM(t) - 2 \sigma(\EFOM), \quad E_{\text{max}}(t) = \EFOM(t) + 2 \sigma(\EFOM),
\end{equation}
with $\sigma(\EFOM)$ denoting the standard deviation of the FOM kinetic energy.
The percentage of time a model is \emph{most accurate} is defined as the
fraction of time steps at which that model achieves the smallest instantaneous
error among all models considered.

\begin{figure}[!ht]
    \centering
    \includegraphics[width=\linewidth]{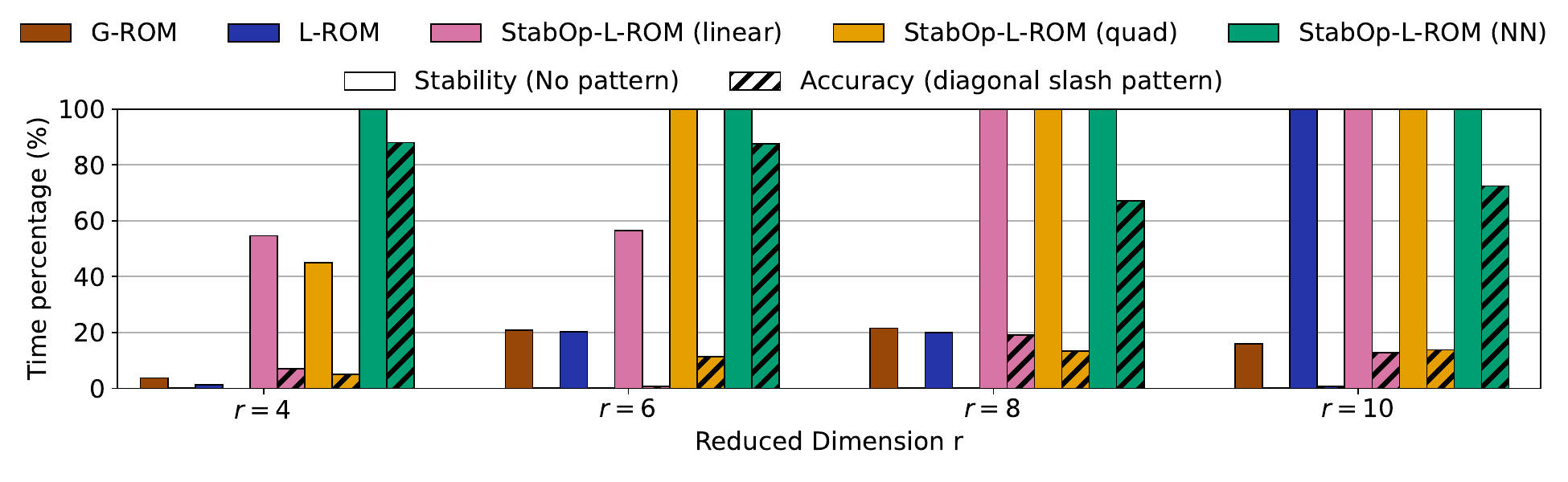}
    \caption{2D flow past a cylinder at $\rm Re=500$. Percentage of time the ROMs are stable (bars with no patterns) and most accurate (bar with diagonal slash patterns).}
    \label{fig:stability-accuracy-cyl}
\end{figure}

For the 2D flow past a cylinder, Fig.~\ref{fig:stability-accuracy-cyl} 
shows that both the G-ROM and the L-ROM exhibit poor performance for $r=4,6$, and $8$. 
In particular, for $r=4$, both models are unstable and never achieve the highest
accuracy among the models considered. For $r=6$ and $r=8$, both models show
similar improvements in stability, but neither of them achieves the highest
accuracy at any time step, and therefore remain the least accurate among the
models considered. For $r=10$, although the L-ROM remains stable throughout the
simulation, 
it remains less accurate than all the {\dlrom}s.

The results in Fig.~\ref{fig:stability-accuracy-cyl} further show that the {\dlrom} (NN) remains stable and is the most
accurate for most of the time (time percentage exceeding $70\%$) for all the ROM 
dimensions $r$ considered. In contrast, the {\dlrom} (linear) and {\dlrom}
(quad) show reduced stability for low 
ROM dimensions, specifically for $r=4$
for both variants, and for $r=6$ for the linear variant.

\begin{figure}[!ht]
    \centering
    \includegraphics[width=\linewidth]{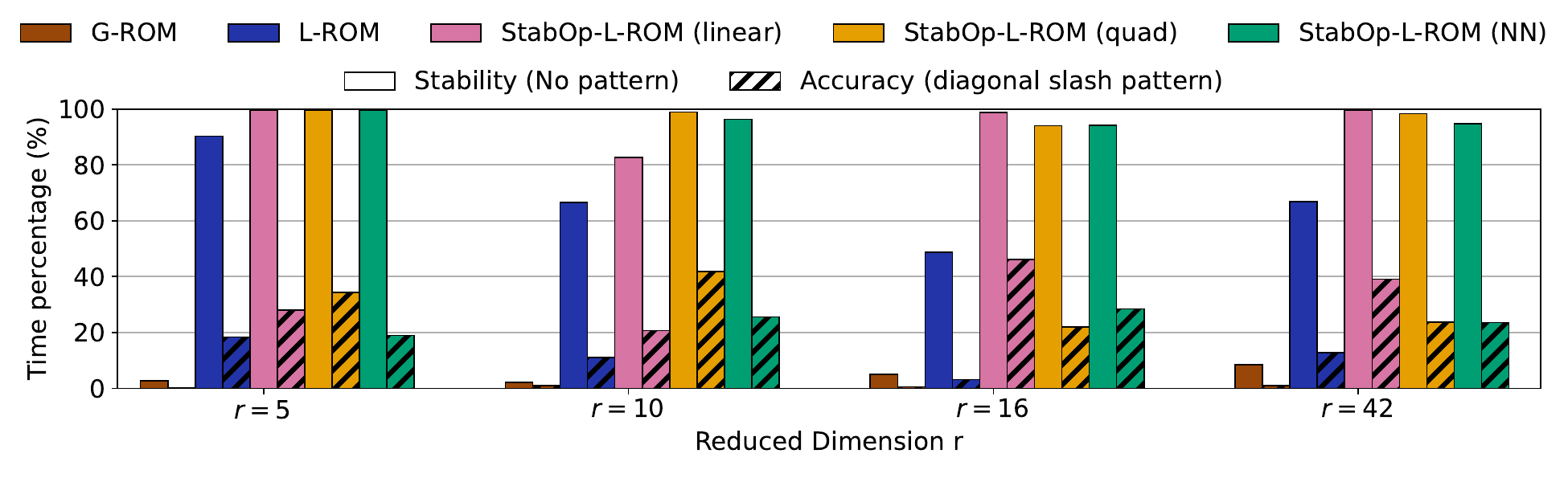}
    \caption{2D lid-driven cavity flow at $\rm Re={10000}$. Percentage of time the ROMs are stable (bars with no patterns) and most accurate (bar with diagonal slash patterns).} 
    \label{fig:stability-accuracy-ldc}
\end{figure}

For the 2D lid-driven cavity, Fig.~\ref{fig:stability-accuracy-ldc} shows that the G-ROM remains unstable and never achieves the highest
accuracy for all ROM dimensions $r$ considered. 
The L-ROM equipped with an optimal filter radius shows a substantial improvement in stability compared to the G-ROM, remaining stable for at least $50\%$ of the simulation time. 
However, despite this improvement, the L-ROM is only the most accurate model for a limited fraction of the simulation time, no more than $20\%$, for all $r$ values.
In contrast, all three {\dlrom}s remain stable for more than $80\%$ of the time across all values of $r$. 
In terms of accuracy, each {\dlrom} achieves the highest accuracy for a similar fraction of the simulation time (roughly between $20\%$ and $40\%$), indicating comparable relative accuracy performance among the StabOp-L-ROM variants and consistent improvement over the G-ROM and L-ROM.

\begin{figure}[!ht]
    \centering
    \includegraphics[width=\linewidth]{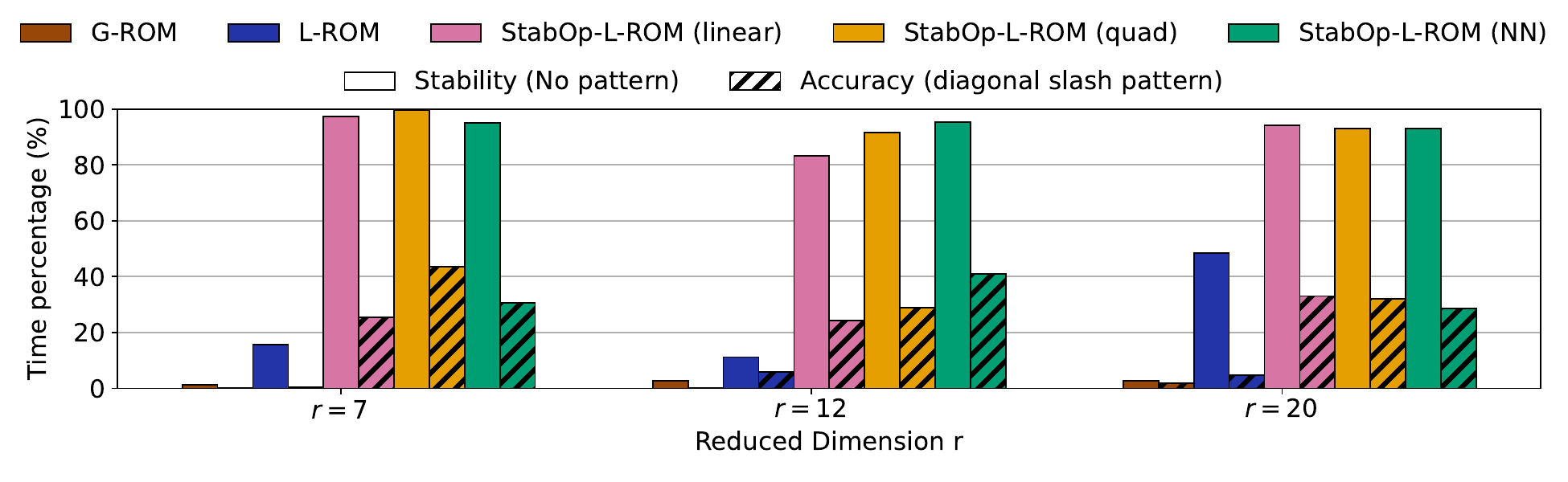}
    \caption{3D flow past a hemisphere at $\rm Re=2200$. Percentage of time the ROMs are stable (bars with no patterns) and most accurate (bar with diagonal slash patterns).}
    \label{fig:stability-accuracy-hemi}
\end{figure}

For the 3D flow past a hemisphere, Fig.~\ref{fig:stability-accuracy-hemi}, 
shows that the G-ROM exhibits poor performance, remaining unstable and
never attaining the highest accuracy for all ROM dimensions $r$ considered.
The L-ROM demonstrates improved stability relative to the G-ROM, particularly at
higher ROM dimensions. However, it achieves the highest accuracy for only a
limited fraction of the simulation time.
In contrast, all three {\dlrom}s remain stable for more than $80\%$ of the time
across all values of $r$. In terms of accuracy, the {\dlrom}s account for most
time steps at which the smallest instantaneous error is attained. Moreover, 
each variant achieves the highest accuracy
for a comparable fraction of the simulation time, indicating similar relative
accuracy performance among the {\dlrom}s and a consistent
improvement over the G-ROM and L-ROM for this test case.

\begin{figure}[!ht]
    \centering
    \includegraphics[width=\linewidth]{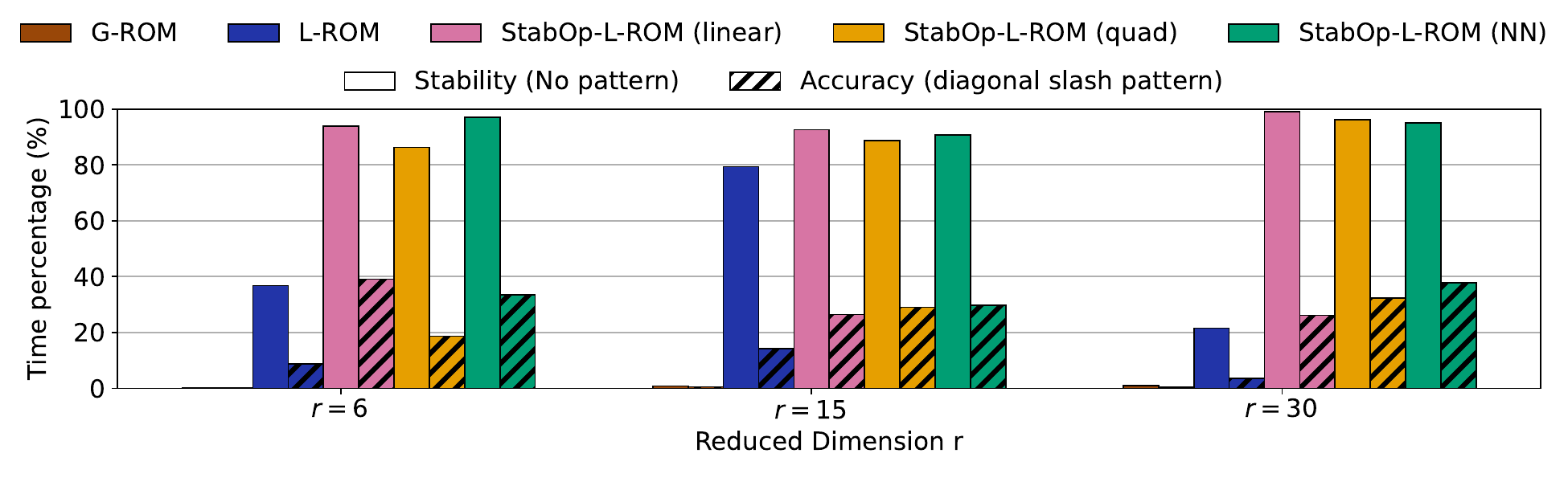}
    \caption{3D minimal channel flow at $\rm Re=5000$. Percentage of time the ROMs are stable (bars with no patterns) and most accurate (bar with diagonal slash patterns).}
    \label{fig:stability-accuracy-mfu}
\end{figure}

The results for the 3D minimal channel flow, shown in Fig.~\ref{fig:stability-accuracy-mfu}, are similar to those observed in the 3D flow past a hemisphere. In particular, the G-ROM again exhibits poor performance, while all three {\dlrom} variants remain stable for most of the simulation time and account for the majority of time steps at which the smallest instantaneous error is achieved. The primary difference is that the L-ROM exhibits improved stability (which could be explained by its over-diffusive character, as illustrated in Figs. \ref{fig:mfu_ke} and \ref{fig:spectrum-mfu}) and a slightly higher accuracy than in the hemisphere case. We note, however, that the L-ROM is still consistently outperformed by the {\dlrom}s.

\begin{figure}[!ht]
    \centering
    \includegraphics[width=\linewidth]{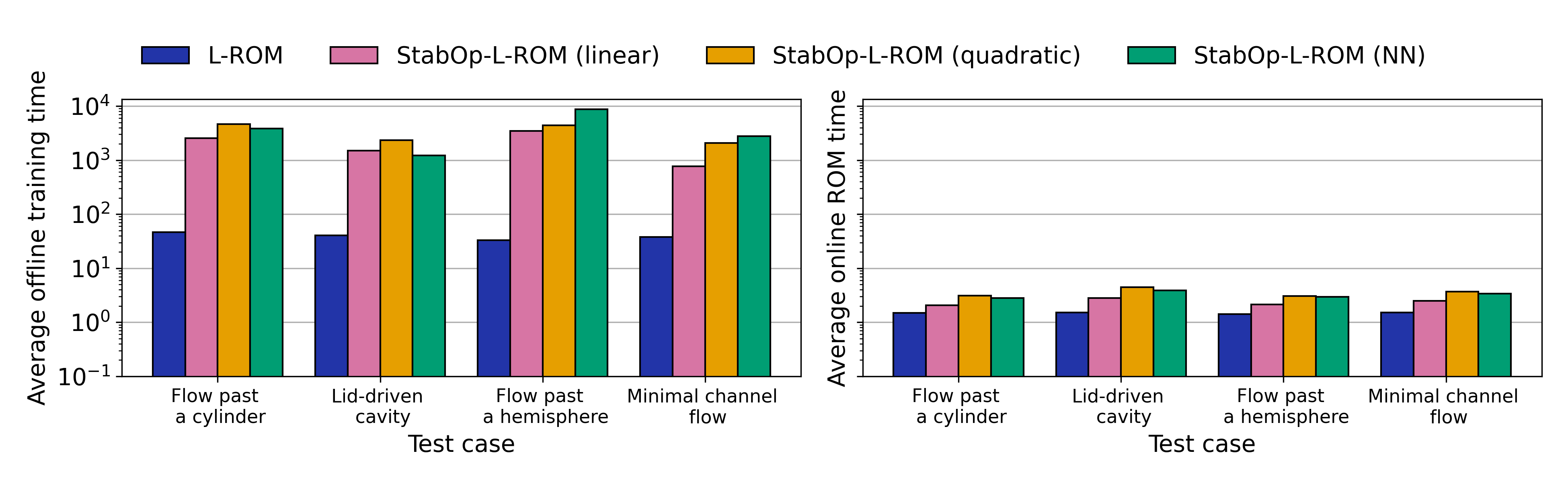}
    \caption{Average computational times needed to perform the L-ROM and StabOp-L-ROMs offline training (left) and the online ROM simulations (right). 
    }
    \label{fig:cpu-times}
\end{figure}

Fig.~\ref{fig:cpu-times} (left) reports the average offline computational time required to train the L-ROM and the {\dlrom}s for each test case. 
For the L-ROM, the reported timings are averaged over all reduced dimensions $r$ considered and correspond to the cumulative time over the filter radius hyperparameter space $\mP_\delta$, reflecting the total offline cost required to identify the optimal $\delta$.
For the {\dlrom}s, the reported timings are averaged over all reduced dimensions 
$r$ and correspond to a single representative choice of hyperparameters. In this case, the filter radius enters only through the initialization of the linear StabOp parameters, while the StabOp is learned via PDE-constrained optimization. Accordingly, timings are reported for a single representative value of the filter radius. The {\dlrom} offline cost is higher than that of the classical L-ROM due to the need to solve a PDE-constrained optimization problem in the construction of the StabOp. 
The offline cost could be further reduced by adopting roll-out training strategies, in which the PDE-constrained optimization is performed over short time horizons rather than the full training interval. This approach avoids the construction of long computational graphs and enables the use of first-order stochastic optimization methods such as Adam, potentially leading to substantial reductions in memory usage and wall-clock time. While such strategies trade global-in-time gradient information for computational efficiency, they have proven effective in training dynamical systems models and represent a promising direction for accelerating the StabOp construction \cite{kim2023generalizable}. 

Fig.~\ref{fig:cpu-times} (right) shows the average online ROM simulation time over the target time interval for each test case. The timings are averaged over all reduced dimensions $r$. 
Despite the higher offline training cost, the {\dlrom}s exhibit only a slightly higher online computational cost than the classical L-ROM, while remaining computationally inexpensive across all test cases.

Direct ROM speedup comparisons with respect to the FOM are not considered because the FOM simulations are executed in parallel and on GPUs, whereas the ROM training procedures are performed serially on CPUs. Here, we report the timings of the FOM simulations to provide the offline snapshot collection cost.
The FOM simulations used to generate the training data are performed using Nek5000 \cite{fischer2008nek5000} for the 2D test cases and NekRS \cite{fischer2022nekrs} for the 3D test cases. The 2D FOM simulations are executed in parallel on CPUs using 8 MPI ranks, requiring approximately 339 seconds for the 2D flow past a cylinder and 4399 seconds for the 2D lid-driven cavity. The 3D FOM simulations are executed on a single NVIDIA TITAN V GPU using NekRS, requiring approximately 1208 seconds for the 3D flow past a hemisphere and 4907 seconds for the 3D minimal channel flow.

\subsection{Is the New StabOp a Spatial Filter?}
    \label{section:numerical-results-spatial-filter}

In this section, we examine whether the StabOp, $\mF$, behaves as a spatial filter. 
To this end, we conduct an experiment where an unphysical G-ROM 
velocity field is used as input to three operators: the new StabOp, the ROM projection (\ref{eq:romproject-weak}), and the classical ROM differential filter (\ref{equation:df-weak}). The three resulting velocity fields are then compared to answer the following two questions: 
(i) {\it Is the StabOp velocity field smoother (e.g., has smoother spatial features) than the input G-ROM velocity field?} If so, we conclude that the new StabOp has the smoothing properties of a ROM spatial filter.
(ii) {\it Is the StabOp velocity field significantly different from the ROM projection and the ROM differential filter velocity fields?}
If so, we conclude that StabOp is different from these classical ROM spatial filters.
This analysis is carried out for three test problems: the 2D lid-driven cavity (Section~\ref{results-ldc}), the 3D flow past a hemisphere (Section~\ref{results-hemi}), and the 3D minimal channel flow (Section~\ref{results-minimal}). 
We note that, while representative results are presented for a single G-ROM solution in each case, the observed behavior of 
the new StabOp is consistent across multiple G-ROM solutions within each test case.

\subsubsection{2D Lid-Driven Cavity}

\begin{figure}[!ht]
    \centering
    \includegraphics[width=1\textwidth]{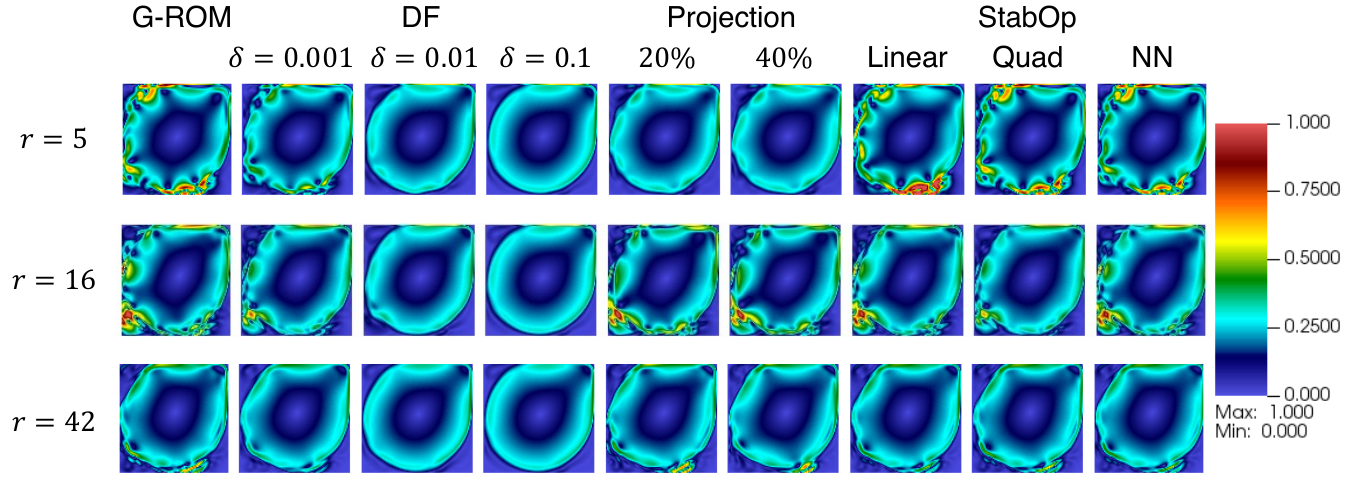}
    
   \caption{2D lid-driven cavity at $\rm Re={10000}$. Velocity magnitudes, where the velocity is obtained by applying the ROM differential filter (DF), the ROM projection, and the three StabOp variants to an unphysical G-ROM solution 
   for $r=5$, $r=16$, and $r=42$.}
   \label{fig:2dldc_filter}
\end{figure}

Figure~\ref{fig:2dldc_filter} shows the velocity fields for $r=5, 16$, and $42$, obtained using the new StabOp with the three model forms (i.e., linear, quadratic, and NN), trained via Algorithm~\ref{alg:d2lrom-training}. These StabOps are the same as those used to produce the kinetic energy predictions shown in Fig.~\ref{fig:2dldc_ke}. Note, however, that the velocity field here differs from those in Fig.~\ref{fig:2dldc_ke}, because in Fig.~\ref{fig:2dldc_filter} the StabOp is applied to a generic G-ROM field, whereas the fields in Fig.~\ref{fig:2dldc_ke} are generated by using the {\dlrom}.  
For comparison, we also include results for the ROM projection with $20\%$ and $40\%$ of the ROM basis functions truncated, and the ROM differential filter with $\delta=0.001$, $0.01$, and $0.1$. 

For $r=5$, none of the three StabOp variants produce  a visibly smoother velocity field, indicating that StabOp does not act as a spatial filter for these $r$ values. However, among the three variants, the linear StabOp yields a velocity field that differs more noticeably from the G-ROM solution than the quadratic and NN variants.
In contrast, for $r=16$ and $r=42$, all three StabOp variants yield velocity fields that are smoother 
than the G-ROM velocity field, suggesting that 
StabOp exhibits filtering behavior. In addition, the StabOp velocity fields remain qualitatively distinct from the velocity fields filtered with the ROM projection and the differential filter. 
This shows that the StabOp filtering is different from the classical ROM projection and ROM differential filtering. 

\begin{figure}[!ht]
    \centering
    \includegraphics[width=1\textwidth]{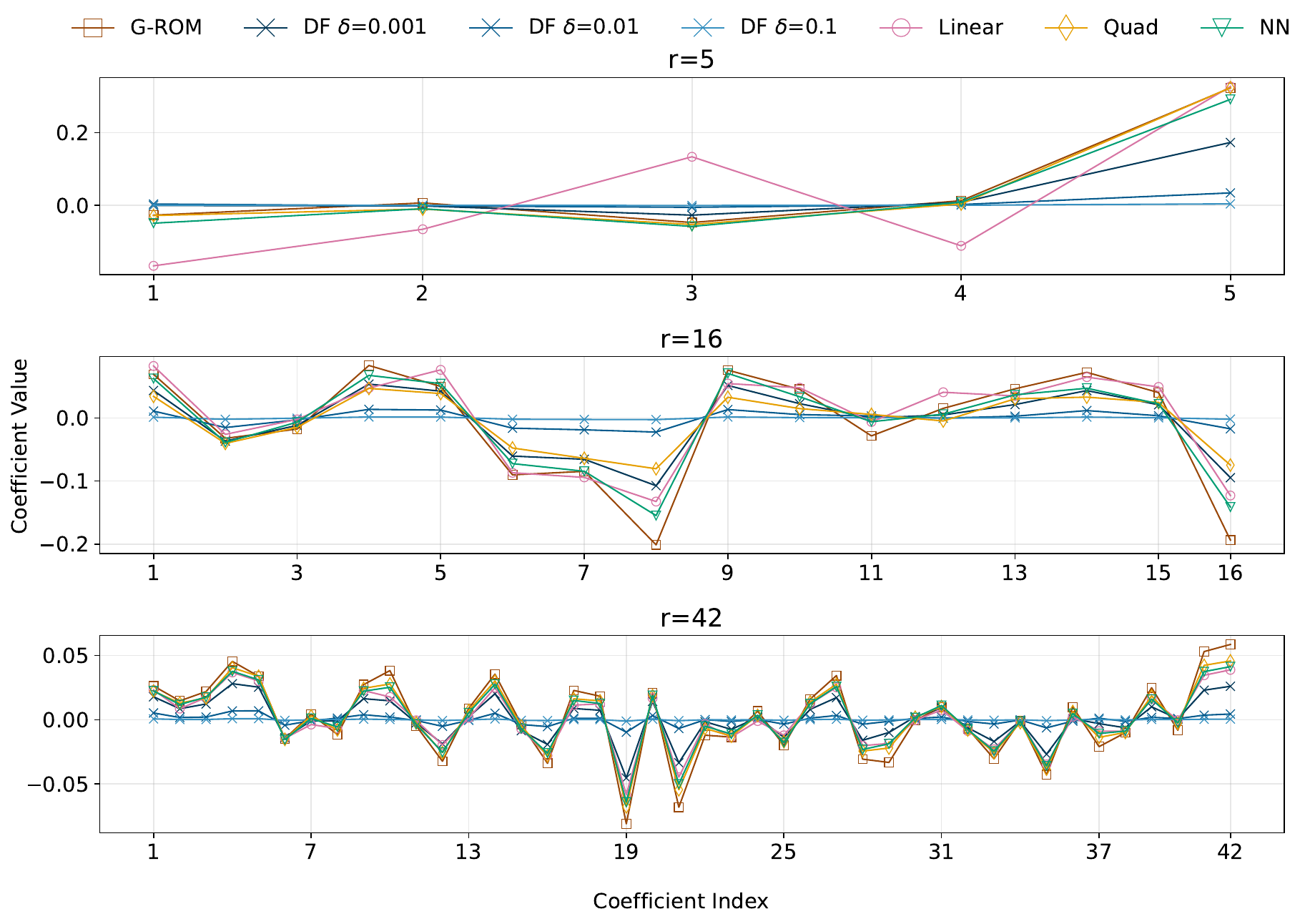}
    
   \caption{2D lid-driven cavity at $\rm Re={10000}$. ROM coefficients obtained by applying the ROM differential filter (DF) and the three StabOp variants 
   to an unphysical G-ROM solution for $r=5$, $r=16$, and $r=42$. 
   }
   \label{fig:2dldc_filter_coef}
\end{figure}

In Fig.~\ref{fig:2dldc_filter_coef}, we plot the ROM coefficients that
correspond to the velocity fields in Fig.~\ref{fig:2dldc_filter}. For clarity,
the ROM projection 
coefficients are not shown, as they coincide with the G-ROM
coefficients for the retained modes and differ only in that the coefficients of the truncated
higher-order modes are set to zero. 
For $r=5$, the ROM coefficients obtained using the StabOp (quadratic) and StabOp (NN) remain close in magnitude to 
the coefficients obtained with the G-ROM. In contrast, the
StabOp (linear) produces noticeably larger coefficient amplitudes across several
modes. This behavior is consistent with the velocity fields shown in
Fig.~\ref{fig:2dldc_filter}, where the StabOp (linear) yields a visibly different
field from the G-ROM solution, while the StabOp (quad) and StabOp (NN) yield fields that are  
similar to the G-ROM field. For $r=16$, the ROM differential filter yields
coefficients with reduced amplitudes across most modes, while the ROM
coefficients obtained using the StabOp exhibit magnitudes and oscillatory
patterns that are broadly comparable to those of the differential filter, but
not identical. This similarity to the ROM differential filter helps explain the
improved smoothness observed in the corresponding velocity fields, whereas the
remaining coefficient differences in the fifth, twelfth, and fifteenth modes
account for the visible discrepancies between the StabOp and the ROM differential filter
fields.
Unlike in the lower-dimensional cases, for $r=42$ the ROM coefficients produced
by the StabOp not only follow the overall trend of the G-ROM coefficients but
are also bounded by them. In particular, the StabOp coefficients generally have
smaller magnitudes than 
the G-ROM coefficients, while remaining larger than those produced by the ROM differential filter. This behavior explains why
the StabOp yields velocity fields that are smoother than the G-ROM 
fields, yet
not as strongly damped as those obtained using the ROM differential filter, and
also why they differ from the ROM projection 
fields, which simply truncate
higher-order modes without modifying the amplitudes of the retained
coefficients.

\subsubsection{3D Flow Past A Hemisphere}

\begin{figure}[!ht]
    \centering
    \includegraphics[width=1.0\textwidth]{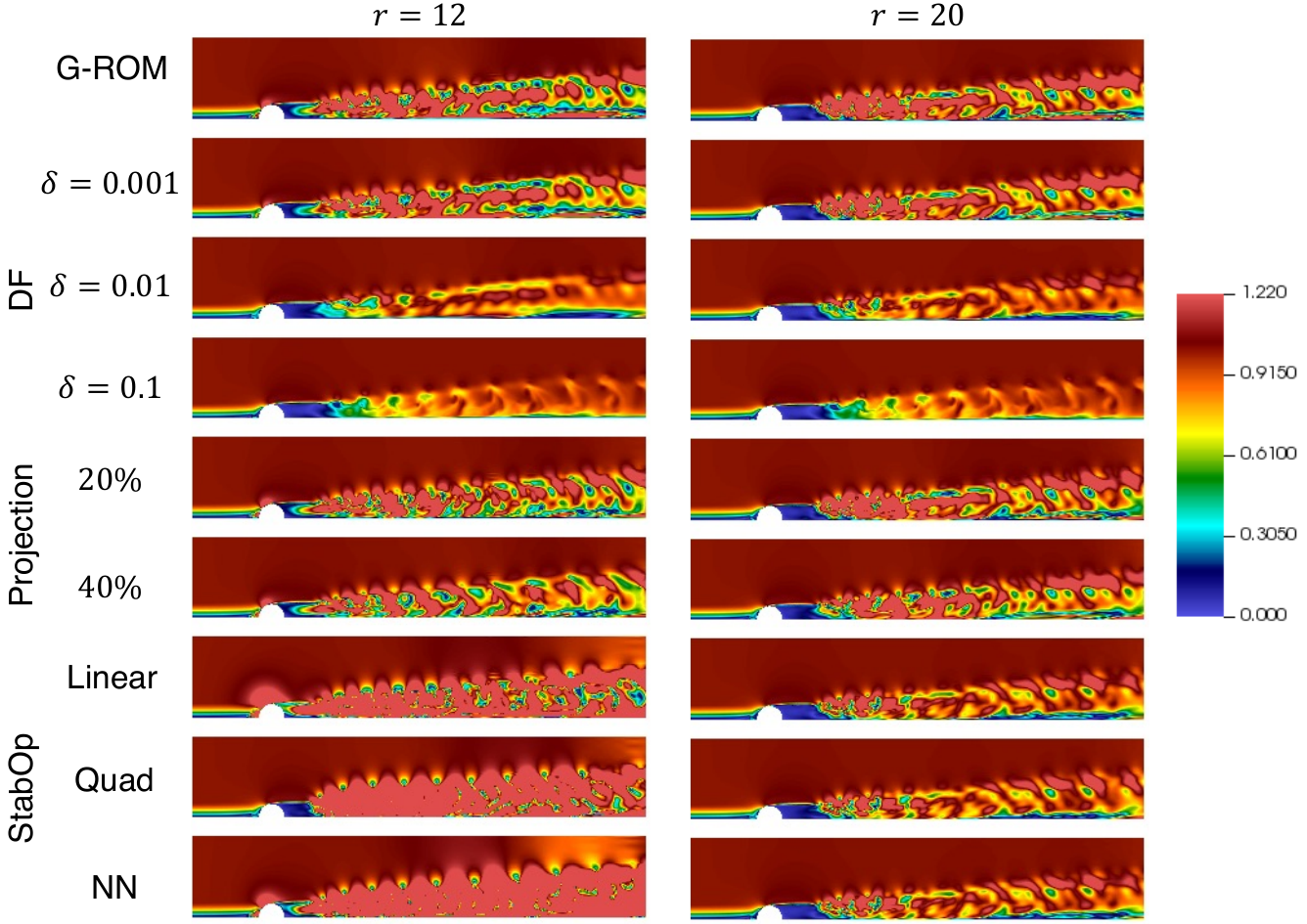}
    
   \caption{3D flow past a hemisphere at $\rm Re={2200}$. Cross section of velocity magnitudes 
   at $y=0$, where the velocity is obtained by applying the ROM differential filter (DF), the ROM projection, and the three StabOp variants 
   to an unphysical G-ROM solution for $r=12$ and $r=20$.}
   \label{fig:hemi_filter}
\end{figure}

Figure~\ref{fig:hemi_filter} shows the velocity fields for $r=12$ and $20$,
obtained using the new StabOp with the three model forms (i.e., linear,
quadratic, and NN), trained via Algorithm~\ref{alg:d2lrom-training}. These
StabOps 
are the same as those used to produce the kinetic energy
predictions shown in Fig.~\ref{fig:hemi_ke}. For comparison, we also include
results for the ROM projection with $20\%$ and $40\%$ of the ROM basis functions
truncated, and the ROM differential filter with $\delta=0.001$, $0.01$, and $0.1$.

For $r=12$, none of the three StabOp variants produce a smoother velocity field, indicating that 
StabOp does not act as a spatial filter. In contrast, for $r=20$, all three StabOp variants yield velocity fields that are noticeably smoother than the G-ROM 
velocity field, suggesting that StabOp exhibits filtering behavior. In addition, the StabOp velocity fields 
remain qualitatively distinct from those obtained with the ROM projection and the ROM differential filter, indicating that StabOp performs a different type of filtering. 

\begin{figure}[!ht]
    \centering
    \includegraphics[width=1\textwidth]{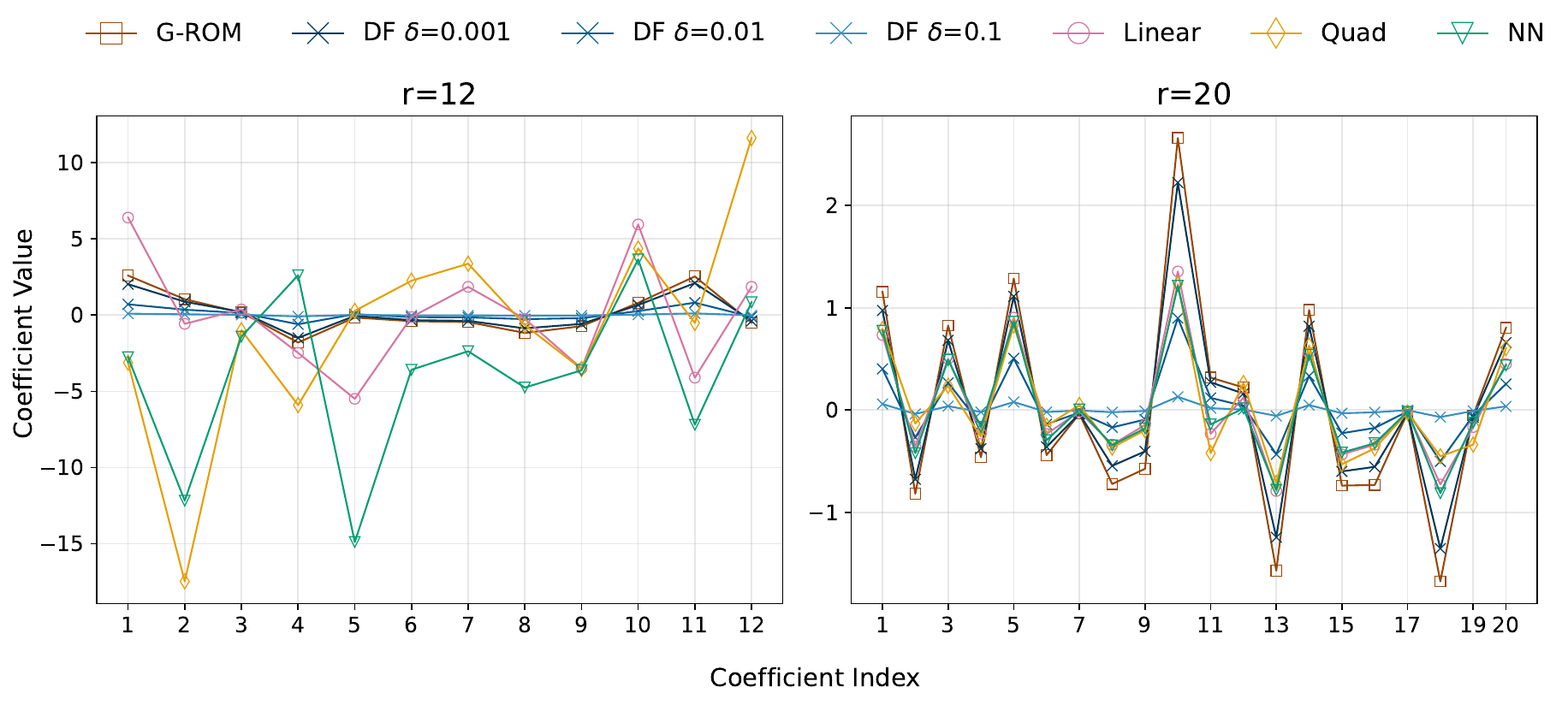}
    
   \caption{3D flow past a hemisphere at $\rm Re={2200}$. ROM coefficients obtained by applying the ROM differential filter (DF) and the three StabOp variants 
   to an unphysical G-ROM solution for $r=12$ and $r=20$.}
   \label{fig:hemi_filter_coef}
\end{figure}

In Fig.~\ref{fig:hemi_filter_coef}, we plot the ROM coefficients that correspond 
to the velocity fields in Fig.~\ref{fig:hemi_filter}. For clarity, the ROM projection 
coefficients are not shown, as they coincide with the G-ROM coefficients for the retained modes and differ only in that the coefficients of the truncated higher-order modes are set to zero. 

For $r=12$, Fig.~\ref{fig:hemi_filter} shows that 
none of the three StabOp variants produce a 
velocity field that is smoother than the input G-ROM field. The ROM coefficient behavior in Fig.~\ref{fig:hemi_filter_coef} is consistent with the velocity field behavior in Fig.~\ref{fig:hemi_filter}.  
In particular, the amplitudes of most ROM coefficients obtained using the StabOp are larger than those of the corresponding G-ROM coefficients, indicating an amplification rather than a damping of modal contributions. 
In contrast, for $r=20$, Fig.~\ref{fig:hemi_filter} shows that all three StabOp variants produce a smoother velocity field. 
The ROM coefficient behavior in Fig.~\ref{fig:hemi_filter_coef} is again consistent with this trend. 
In particular, the ROM coefficients obtained using the StabOp exhibit magnitudes and oscillatory patterns that are comparable to the ROM differential filter ROM coefficients, 
{but not identical.} 
This similarity to the ROM differential filter helps explain the improved smoothness observed in the corresponding velocity fields, whereas the remaining coefficient differences, such as the amplification of the eleventh and nineteenth modes, account for the visible discrepancies between the StabOp and ROM differential filter fields.

\subsubsection{3D Minimal Channel Flow}

\begin{figure}[!ht]
    \centering
    \includegraphics[width=1.0\textwidth]{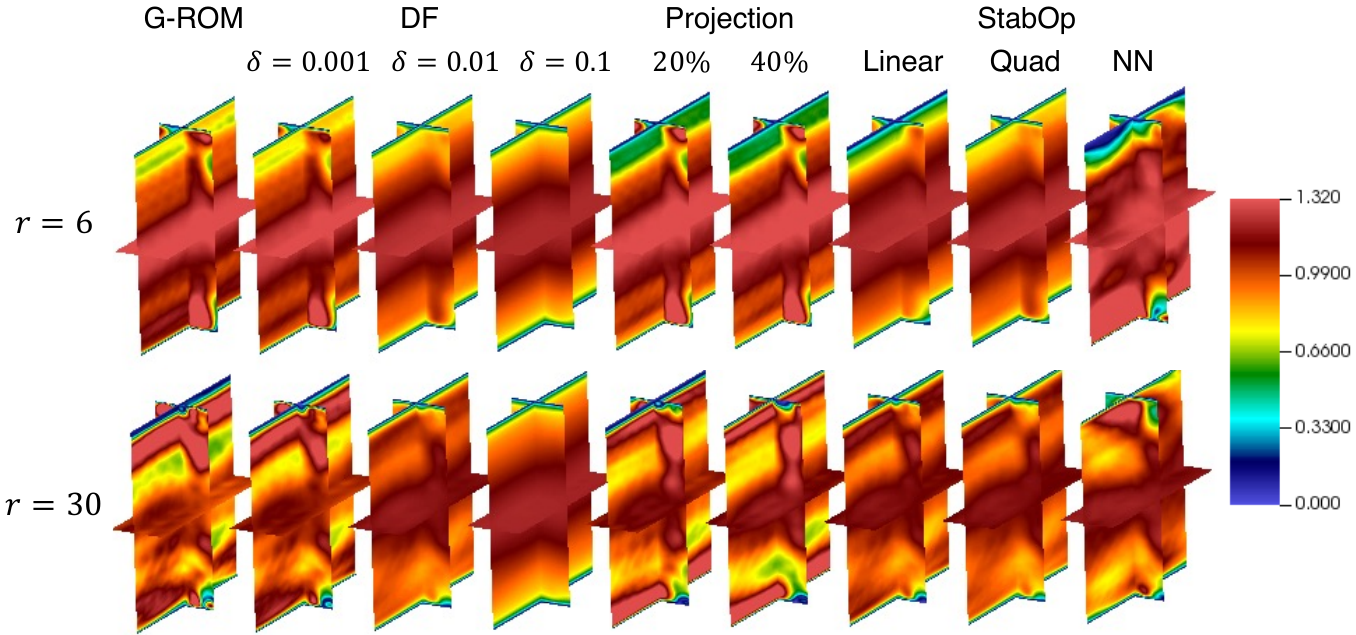}
    
   \caption{3D minimal channel flow at $\rm Re={5000}$. Velocity magnitudes, where the velocity is obtained by applying the ROM differential filter (DF), the ROM projection, and the three StabOp variants 
   to an unphysical G-ROM solution for $r=6$ and $r=30$.}
   \label{fig:mfu_filter}
\end{figure}

Figure~\ref{fig:mfu_filter} shows the velocity fields for $r=6$ and $30$,
obtained using the new StabOp with the three model forms (i.e., linear,
quadratic, and NN), trained via Algorithm~\ref{alg:d2lrom-training}. These
StabOps 
are the same as those used to produce the kinetic energy
predictions shown in Fig.~\ref{fig:mfu_ke}. For comparison, we also include
results for the ROM projection with $20\%$ and $40\%$ of the ROM basis functions
truncated, and the ROM differential filter with $\delta=0.001$, $0.01$, and $0.1$.

For both 
ROM dimensions, the 
velocity fields produced by 
the StabOp are noticeably smoother than the G-ROM 
fields, indicating that 
StabOp exhibits a filtering behavior. 
However, clear differences are observed among the three StabOp variants. The
linear and quadratic variants produce velocity fields that are
qualitatively similar to each other, whereas the NN model yields velocity fields
with distinct structural characteristics. For example, it introduces additional
features in both the near-wall and central regions of the channel.
We also note that the StabOp fields 
are qualitatively different from the ROM differential filter fields. 
In particular, the ROM differential filter yields smoother structures near the channel walls, whereas the 
StabOp fields retain more features in those regions. In addition, in the central region of the channel, the 
StabOp 
fields retain richer flow structures compared to the ROM differential filter fields, suggesting that 
StabOp applies a less aggressive and potentially more physically meaningful form of filtering. 
In contrast, the ROM projection yields velocity fields that are less smooth than both the ROM differential filter and 
StabOp fields.

\begin{figure}[!ht]
    \centering
    \includegraphics[width=1\textwidth]{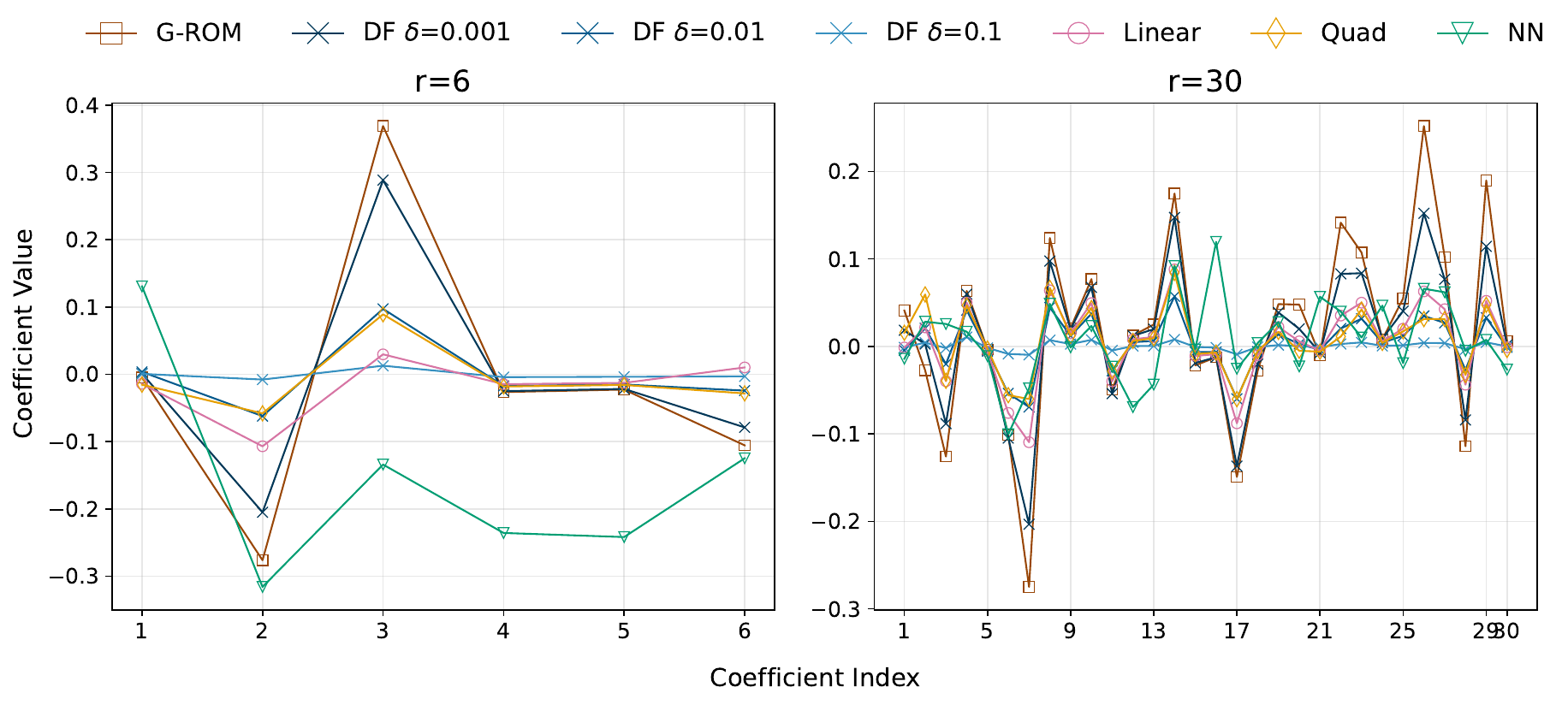}
    
   \caption{3D minimal channel flow at $\rm Re={5000}$. ROM coefficients obtained by applying the ROM differential filter (DF) and the three StabOp variants to an unphysical G-ROM solution for $r=6$ and $r=30$.}
   \label{fig:mfu_coef}
\end{figure}

In Fig.~\ref{fig:mfu_coef}, we plot the ROM coefficients that correspond
to the velocity fields in Fig.~\ref{fig:mfu_filter}. For clarity, the ROM
projection coefficients are not shown, as they coincide with the G-ROM coefficients
for the retained modes and differ only in that the coefficients of the truncated higher-order modes
are set to zero. For $r=6$, clear differences emerge among the StabOp variants
at the coefficient level. The StabOp (linear) and StabOp (quad) yield
coefficients with smaller amplitudes than those of the G-ROM, although they do
not exactly follow the trend of the ROM differential
filter coefficients. This behavior explains why the corresponding velocity fields
appear smoother than the G-ROM solution, while still remaining distinct from the
differential filter fields. In contrast, the StabOp (NN) exhibits 
a different coefficient behavior. Most of the modes are amplified relative to both the
G-ROM and the other StabOp variants. This amplification is consistent with the
significantly altered velocity fields observed for the StabOp (NN). For $r=30$, the
overall behavior of coefficients associated with the StabOp (linear) and StabOp
(quad) continues to follow the trend of the ROM differential filter, while
remaining distinct from it. The coefficients associated with the StabOp (NN) deviate
from this trend, with several modes exhibiting amplification compared with the other
StabOp variants and the ROM differential filter.

\section{Conclusions}
    \label{section:conclusions}

Spatial filtering has long been a cornerstone in the development of numerical methods for turbulent flow simulation. In classical 
LES, a wide range of spatial filters have been employed to construct effective turbulence models. 
More recently, spatial filters have also been adapted to increase the ROM stability and accuracy, leading to a wide variety of ROM stabilizations and closures. 
Finally, spatial filtering has also been leveraged to increase the stability of neural-network-based models. 
The success of numerous filter-based closures and stabilizations in LES and ROMs demonstrate that spatial filters are a highly effective approach for modeling turbulent flows. 
However, that spatial filters are not perfect.
Let us consider, for example, the following practical scenario:
Given a realistic coarse resolution and a stabilization strategy, which filter should be chosen from the wide array of choices?
Furthermore, once the filter is chosen, how do we choose the filter parameters, e.g., the filter radius?
The answers to these questions are critical for the success of the stabilization strategy:
The right choices can yield accurate results, whereas the wrong choices can yield 
highly inaccurate results.
Thus, the following natural question arises: 
{\it Are classical spatial filters truly the best tool for constructing closures and stabilizations?} 

In this paper, we 
demonstrate that spatial filters are \underline{\it not} always the best tool for constructing closures and stabilizations.  
Specifically, for ROMs and a particular 
stabilization strategy (i.e., the Leray ROM (L-ROM)), we show that we can construct a novel {\it data-driven 
stabilization operator (StabOp)} that yields more accurate results than the classical ROM differential filter in the following sense:
By replacing the ROM differential filter with the new StabOp in the Leray ROM stabilization, we obtain a new ROM stabilization, denoted StabOp-L-ROM.
To construct the novel 
StabOp, we first postulate a model form, 
which could be linear, quadratic, or nonlinear (neural network based).
Then, we use the available data and solve a {\it PDE-constrained optimization} problem to find the model parameters that yield the most accurate StabOp-L-ROM. 
Specifically, for a given QoI (e.g., the kinetic energy), we minimize a loss function that quantifies the QoI difference between the StabOp-L-ROM and the FOM, and use the StabOp-L-ROM as the PDE constraint for the minimization problem. 

Our new data-driven 
stabilization operator, {\it StabOp, 
is a fundamental departure from classical filter-based stabilization}: 
Rather than relying on the long-standing principle that spatial filtering improves model stability and accuracy, StabOp 
instead learns a stabilization operator directly from data, with the goal of maximizing QoI's accuracy for the 
given stabilization model.
We also note that the new StabOp is a {\it model-centric} stabilization operator:
To build StabOp, we constrain the minimization of the loss function to satisfy the PDE associated with the specific stabilization model (i.e., L-ROM in our case).
This strategy enables us to construct stabilization operators that are optimal for the chosen stabilization (i.e., L-ROM).  

To investigate the new StabOp, we perform two sets of numerical experiments.
In the first set, we compare the new StabOp-L-ROM with the standard L-ROM equipped with the classical ROM differential filter.
For a fair comparison, we choose an optimal filter radius for the differential filter in L-ROM.
As test problems, we use the following  
convection-dominated flows: 
2D flow past a cylinder at $\rm Re=500$, 
lid-driven cavity at $\rm Re=10000$, 3D flow past a hemisphere at $\rm Re=2200$, and minimal channel flow at $\rm Re=5000$. 
Our first set of numerical tests show that the new {\it StabOp-L-ROM can be orders-of-magnitude more accurate than 
the classical L-ROM} with the optimal filter radius: 
StabOp-L-ROM yields large improvements in accuracy over the standard L-ROM with optimized parameters for one test case (see Fig.~\ref{fig:mse_ldc}), and one or even more orders of magnitude improvements for the remaining test cases (see Figs.~\ref{fig:mse_cyl}, \ref{fig:mse_hemi}, and \ref{fig:mse_mfu}).
In the second set of numerical tests, we compare the new StabOp with two classical ROM filters:
the ROM differential filter and the ROM projection.
Specifically, we consider an inaccurate velocity field generated by an under-resolved G-ROM as input, and apply the new StabOp, the ROM differential filter, and the ROM projection to it.
We use the same test problems as in the first numerical experiment, except the 2D flow past a cylinder at $\rm Re=500$. 
Our second set of numerical tests show that  StabOp smooths the input field, but its smoothing mechanism is different from that of the ROM differential filter and ROM projection.
These results clearly show that the new {\it StabOp is different from classical ROM spatial filters}.

These first steps in the assessment of the novel StabOp 
clearly show that it can significantly improve the accuracy of classical filter-based stabilizations. 
There are, however, several research directions that should be pursued to gain a better understanding of the new StabOp's potential and possible limitations.
The first step should probably be the investigation of the new 
StabOp in conjunction with other 
ROM stabilizations, e.g., the evolve-filter-relax ROM and the time relaxation ROM.
Furthermore, StabOp could also be used to develop new ROM closures.
Another important research direction is the StabOp's extension from the ROM setting to the FOM realm. 
Finally, one should also develop a numerical analysis framework for the new 
StabOp that is similar to that for 
classical filter-based stabilizations and closures. 

\section*{CRediT Authorship Contribution Statement}

\textbf{Ping-Hsuan Tsai:}
Writing - original draft, Writing - review and editing, Investigation, Conceptualization, Validation, Visualization, Methodology, Software, Formal Analysis;
\textbf{Anna Ivagnes:}
Writing - original draft, Writing - review and editing, Conceptualization, Validation, Methodology, Software, Formal Analysis;
\textbf{Annalisa Quaini:} 
Writing - review and editing, Conceptualization, Methodology, Supervision;
\textbf{Traian Iliescu:} 
Writing - original draft, Writing - review and editing, Conceptualization, Methodology, Supervision;
\textbf{Gianluigi Rozza:} 
Writing - review and editing, Conceptualization, Supervision.

\section*{Acknowledgments}
AI and GR acknowledge the support provided by the European Union - NextGenerationEU, in the framework of the iNEST - Interconnected Nord-Est Innovation Ecosystem (iNEST ECS00000043 – CUP G93C22000610007) consortium and its CC5 Young Researchers initiative. The authors would  also like to acknowledge INdAM-GNCS for its support and MUR PRIN 2022 project FAROM.

\bibliographystyle{plain}
\bibliography{traian,pht}

\end{document}